\documentclass[aip,jcp,amsmath,amssymb,floatfix,reprint,citeautoscript]{revtex4-2}

\usepackage[utf8]{inputenc}
\usepackage{graphicx}
\usepackage{wrapfig}
\usepackage{bibunits}
\usepackage[colorlinks,allcolors=black,citecolor=blue,urlcolor=blue]{hyperref}
\usepackage[mathlines]{lineno}
\usepackage{chemformula}
\usepackage{chemfig}
\usepackage{physics}
\usepackage[debrief]{silence}
\usepackage{siunitx}

\graphicspath{{figures/}}

\defaultbibliography{paper,revtex-control}
\defaultbibliographystyle{aipnum4-2}

\begin{document}

\def\mytitle{Reducing the cost of neural network potential generation for reactive molecular systems}
\title{\mytitle}

\author{Krystof Brezina}

\author{Hubert Beck}

\author{Ondrej Marsalek}
\email{ondrej.marsalek@mff.cuni.cz}

\affiliation{
Charles University, Faculty of Mathematics and Physics, Ke Karlovu 3, 121 16 Prague 2, Czech Republic 
}

\date{\today}

\begin{abstract}

\setlength\intextsep{0pt}
\begin{wrapfigure}{r}{0.4\textwidth}
  \hspace{-1.8cm}
  \includegraphics[width=0.4\textwidth]{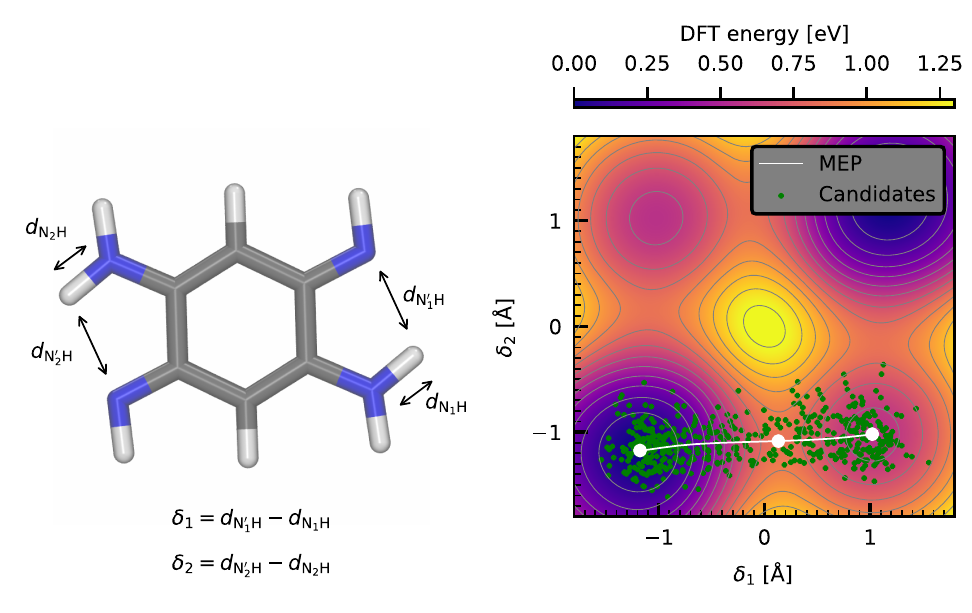}
\end{wrapfigure}

Although machine-learning potentials have recently had substantial impact on molecular simulations, the construction of a robust training set can still become a limiting factor, especially due to the requirement of a reference \textit{ab initio} simulation that covers all the relevant geometries of the system.
Recognizing that this can be prohibitive for certain systems, we develop the method of transition tube sampling that mitigates the computational cost of training set and model generation.
In this approach, we generate classical or quantum thermal geometries around a transition path describing a conformational change or a chemical reaction using only a sparse set of local normal mode expansions along this path and select from these geometries by an active learning protocol.
This yields a training set with geometries that characterize the whole transition without the need for a costly reference trajectory.
The performance of the method is evaluated on different molecular systems with the complexity of the potential energy landscape increasing from a single minimum to a double proton-transfer reaction with high barriers.
Our results show that the method leads to training sets that give rise to models applicable in classical and path integral simulations alike that are on par with those based directly on \textit{ab initio} calculations while providing the computational speed-up we have come to expect from machine-learning potentials.

\end{abstract}

{\maketitle}

\begin{bibunit}

\nocite{revtex-control}

\section{Introduction}

Owing to the detailed atomistic insight into the structure and dynamics of molecular systems and materials, the relevance of computer simulations of molecular dynamics (MD) in current research is undeniable.
MD simulations represent a valuable analytic and predictive tool in multiple fields of both basic and applied research including physical chemistry, material science, or drug design~\cite{Tian2008/10.1039/b703897f,DeVivo2016/10.1021/acs.jmedchem.5b01684,Hollingsworth2018/10.1016/J.NEURON.2018.08.011,Deringer2019/10.1002/adma.201902765,Yao2022/10.1021/ACS.CHEMREV.1C00904}.
They also provide a way to explain and corroborate experimental data that might be difficult to interpret otherwise.
For many systems of interest, MD simulations can be routinely performed under the Born--Oppenheimer approximation in the electronic ground state, with the nuclei being treated either classically or quantum-mechanically within the imaginary-time path integral formalism.
This makes the choice of the potential energy surface (PES) a key decision that determines the accuracy of the resulting simulation.
Out of the available options, \textit{ab initio} molecular dynamics~\cite{Marx2009} (AIMD) is a state-of-the-art methodology that relies on a full, on-the-fly quantum electronic structure calculation~\cite{Szabo1982,Parr1989} at every step of the simulation to evaluate the potential energy and forces.
This is most commonly performed at the level of density functional theory~\cite{Hohenberg1964/10.1103/PhysRev.136.B864,Kohn1965/10.1103/PhysRev.140.A1133,Parr1989} (DFT), which provides correlated electronic energies at a computational cost accessible in practice, but for smaller systems, the use of correlated wave function methods is feasible as well.~\cite{Marsalek2014/10.1021/ct400911m,Spura2015/10.1039/C4CP05192K,Kapil2016/10.1063/1.4941091}
In any case, the computational cost of AIMD simulations is typically large --- especially so for advanced hybrid DFT functionals in the condensed phase --- and can easily become prohibitive in the light of the ever-growing demand for larger time and length scales of the relevant simulations.

This issue can be mitigated by the recent development of the so-called machine learning potentials (MLPs)~\cite{Behler2011/10.1039/C1CP21668F,Friederich2021/10.1038/s41563-020-0777-6}.
These use various machine learning approaches~\cite{Behler2007/10.1103/PhysRevLett.98.146401,Bartok2010/10.1103/PhysRevLett.104.136403,Smith2017/10.1039/C6SC05720A,Schutt2017/10.48550/arXiv.1706.08566,Batzner2022/10.1038/s41467-022-29939-5} to faithfully approximate the desired \textit{ab initio} PES by training on a reference data set consisting of a relatively modest number of \textit{ab initio} geometries and their corresponding energies and, optionally, forces.
As such, they indeed combine the best of the two worlds: they are able to maintain the accuracy of the parent \textit{ab initio} method, but they also circumvent the need for explicit electronic structure calculations at each step of the MD simulation.
Thus, they evaluate the potential energy and forces at a significantly reduced computational cost~\cite{Dral2020/10.1021/acs.jpclett.9b03664}.
One particular flavor of MLPs of major practical importance is represented by neural network potentials (NNPs), which rely on artificial neural networks combined with a set of appropriate atomic descriptors to accurately represent the molecular geometry-to-energy relationship, including all its symmetries.~\cite{Behler2007/10.1103/PhysRevLett.98.146401,Behler2011/10.1063/1.3553717}
NNPs have repeatedly proved their worth in modeling a plethora of various molecular systems ranging from liquids and solutions to interfaces and solids.~\cite{Natarajan2016/10.1039/c6cp05711j,Schutt2018/10.1063/1.5019779,Sivaraman2020/10.1038/s41524-020-00367-7,Schran2021/10.1073/PNAS.2110077118,Batzner2022/10.1038/s41467-022-29939-5}
Our recent work focusing on NNPs~\cite{Schran2020/10.1063/5.0016004,Schran2021/10.1073/PNAS.2110077118} shows that rather than using a single NNP to represent the PES, it is advantageous to build a model as a committee~\cite{Hansen1990/10.1109/34.58871} of NNPs (C-NNP) that comprises a small number of NNPs, each trained individually to a subset of the main training set.
The advantage is twofold: first, the energy prediction obtained as the committee average is known to be a better approximation of the \textit{ab initio} energy than the estimates of the individual members~\cite{Gastegger2017/10.1039/c7sc02267k,Schran2020/10.1021/acs.jctc.9b00805}.
Second, the committee disagreement~\cite{Krogh1995/10.1.1.37.8876}, represented by the standard deviation of the individual member estimates of energies or forces, serves as a valuable indicator of prediction reliability and can be used to monitor and optionally ensure the stability of the simulation~\cite{Schran2020/10.1063/5.0016004}.
Crucially, this disagreement can be used as the key ingredient of the active learning process called query by committee~\cite{Seung1992/10.1145/130385.130417} (QbC) that systematically builds the training set in a data-driven way~\cite{Krogh1995/10.1.1.37.8876,Schran2020/10.1063/5.0016004,Chen2020/10.1039/c9ra09935b}.

An accurate and stable NNP can only be obtained on a foundation of robust, high-quality training data.
This is typically based on a reference AIMD trajectory, from which geometries are selected for the training set, together with the corresponding energies and forces.
However, the trajectory is highly correlated in time and thus most of the expensive AIMD data does not contribute useful information for the training of the model.
This selection has been approached in different ways from random sampling and manual selection to more data-driven procedures~\cite{Morawietz2016/10.1073/pnas.1602375113,Smith2018/10.1063/1.5023802,Musil2018/10.1039/c7sc04665k,Schran2020/10.1063/5.0016004,Vandermause2020/10.1038/s41524-020-0283-z,Sivaraman2020/10.1038/s41524-020-00367-7,Wang2020/10.1039/d0cc03512b,Miksch2021/10.1088/2632-2153/abfd96,Schwalbe-Koda2021/10.1038/s41467-021-25342-8,Vandenhaute2023/10.1038/s41524-023-00969-x}, with QbC being a particularly efficient method.
QbC considers a set of candidate structures, in this case the whole AIMD trajectory, and iteratively builds up the training set.
It starts by training a C-NNP on a very small set of initial configurations and using its disagreement to screen the candidate configurations for those with the most uncertain prediction.
A small number of these configurations are then added to the training set, a new C-NNP is trained, and the process is iteratively repeated until some convergence criteria are met.
In comparison to random selection, this approach is known to generate more compact training sets that give rise to robust models of similar accuracy~\cite{Smith2018/10.1063/1.5023802,Vandermause2020/10.1038/s41524-020-0283-z}.
Even though the initial AIMD trajectory is typically the most expensive part of the procedure, numerous successful MLPs have been generated on top of reasonably short AIMD simulations.~\cite{Schran2021/10.1073/PNAS.2110077118}

However, for many purposes, this process involving AIMD is still too expensive to be practical.
For instance, the requirements on a high-level electronic structure method can raise the computational demands above a reasonable threshold.
One might also be interested in a system that features rare events, such as chemical or conformational changes, which will happen quickly and occur infrequently or not at all in a direct AIMD simulation.
In turn, these crucial configurations are underrepresented in the set of candidates and enhanced sampling simulations would be required in order to construct a robust training set, which typically raises the computational cost further by one or more orders of magnitude.

In case such a situation occurs, one needs to adhere to an approximate method of candidate generation that relieves some of the computational expenses while maintaining the quality of the resulting candidate set.
For simple systems with a single potential energy minimum, the solution is fairly straightforward.
In this case, one can benefit from random sampling of displacements in the directions of a fixed set of normal modes to obtain a set of distorted configurations.
This approach, sometimes called normal mode sampling (NMS) in the literature~\cite{Smith2017/10.1039/C6SC05720A}, avoids the cost of a full AIMD simulation by replacing it with a more manageable combination of a Hessian matrix evaluation and a number of single-point electronic structure calculations for the generated geometries.
The sampling of the known normal mode distribution itself yields uncorrelated samples by definition and requires no \textit{ab initio} calculations, therefore its cost is negligible.
Various versions of this approach were successfully used to generate structures for the training of MLPs.
Using a scaled uniform random sampling of the normal modes, the method was used to obtain auxiliary structures used in model validation~\cite{Rupp2015/10.1021/ACS.JPCLETT.5B01456} 
and with approximate thermal distortions in NNP training set generation around configurational minima~\cite{Smith2017/10.1039/C6SC05720A} as well as to construct an NNP model for a gas-phase ammonia molecule.~\cite{Schwalbe-Koda2021/10.1038/s41467-021-25342-8}
Clearly, the utility of NMS is limited when the harmonic approximation becomes insufficient.
This can be the case if individual modes are strongly anharmonic or coupled, or if the system features conformational changes or reactions, where multiple local minima come into play.

In this work, we propose transition tube sampling (TTS), a robust and general approach to the generation of training sets and models that are able to accurately describe processes that feature transitions over potential energy barriers, which includes both conformational flexibility and chemical reactivity.
We achieve this by generating thermally distorted candidate geometries along a reaction pathway with the help of multiple normal mode expansions and screening these candidates using QbC.
The role of the minimum geometry in NMS is taken by the minimum energy path (MEP) that describes the course of the reaction through configuration space.
Local harmonic expansions are performed in a small number of relevant configurations along the MEP and physically relevant candidate configurations are generated with uniform distribution along the MEP and with classical or quantum thermal weights in all perpendicular directions based on one of the sets of normal modes.
An arbitrary number of these candidate configurations can be generated at a negligible computational cost and submitted to the QbC process, which selects the most important ones to have \textit{ab initio} calculations performed and to be included in the training set.
This results in compact and robust training sets and models that maintain consistent accuracy along the reaction path, making them suitable for MD simulations of the reactive process, including enhanced sampling simulations, while no AIMD trajectories are required as part of this process.
We test this method on three different molecules in the gas phase to illustrate its capabilities.

The rest of the paper is organized as follows.
In Section~\ref{sec:theory}, we begin by formalizing thermal NMS, which samples the exact classical or quantum canonical distribution under the harmonic approximation.
With the obtained framework, we then proceed to introduce the MEP into the picture and describe the technical details of TTS.
In Section~\ref{sec:computational-details}, we describe how we used TTS to create C-NNP models, the simulations performed with these models, and other related computational details.
In Section~\ref{sec:results}, we apply this approach to three different gas-phase systems of increasing complexity represented by the molecules of benzene, malonaldehyde and 2,5-diaminonenzoquinone-1,4-diimine (DABQDI) and discuss its successes and possible pitfalls.
Section~\ref{sec:conclusions} concludes the paper and offers outlooks concerning the generalization and the limitations of the method beyond the gas phase.

\section{Theory}
\label{sec:theory}

In this section, we discuss the theoretical basis of the TTS method.
In this approach, we rely on the harmonic approximation and the vibrational normal mode formalism to obtain \textit{ab initio} training data for the construction of C-NNPs for reactive systems without the need to run expensive sampling simulations, such as full AIMD. 
First, we present the simple key idea behind NMS which relies on the harmonic approximation to describe the underlying PES and thus is expected to work well for systems that are close to harmonic around a single given minimum geometry at the temperature of interest.
Clearly, this does not yield a flexible and general method, since the harmonic approximation is readily challenged by many realistic systems, notably those that exhibit more pronounced configurational flexibility or chemical reactivity.
Therefore, we propose a more general approach to sampling candidate geometries based on NMS which is applicable even to systems described by multiple minima separated by barriers.
This is achieved by using the harmonic expansion locally along an MEP in a way that eventually generates a balanced training set.

\subsection{NMS for thermal sampling around minimum geometries}

To open the discussion of the theory behind TTS, we first turn our attention to the simple case represented by a PES with a single minimum geometry $\mathbf{R}_0$ on which the nuclear motion is described by classical mechanics.
Assuming a reasonable extent of validity of the harmonic approximation to capture the thermally accessible potential energy landscape, the classical thermal probability density $\rho_\mathrm{c}$ at temperature $T$ is approximated by
\begin{equation}
    \rho_\mathrm{c}(\Omega_1, \dots, \Omega_{N_\mathrm{int}})
    \propto
    \prod_{i=1}^{N_\mathrm{int}} \exp\left(-\frac{1}{2} \beta \omega_i^2 \Omega_i^2\right).
    \label{eq:boltzmann-cl}
\end{equation}
In this expression, $\omega_i$ and $\Omega_i$ denote, respectively, the natural frequency and the normal coordinate corresponding to the $i$-th normalized vibrational normal mode vector $\mathbf{\Omega}_i$, and $\beta$ is the inverse temperature equal to $1 / k_\mathrm{B}T$ (with $k_\mathrm{B}$ representing the Boltzmann constant).
$N_\mathrm{int}$ is the total number of internal degrees of freedom of the species, typically $3N-6$ for $N$ atoms.
Hence, in the harmonic approximation, the thermal density is described as a multivariate, yet uncoupled normal distribution where each $i$-th orthogonal degree of freedom has the standard deviation of $\sigma_i = 1 / \sqrt{\beta\omega_i^2}$.

As such, it is straightforward to generate completely uncorrelated thermal geometries $\mathbf{R}$ by distorting the minimum geometry $\mathbf{R}_0$ independently in the direction of each of the normal mode vectors.
The appropriate magnitude of the distortions is given by a randomly generated value of the corresponding normal coordinate $\Omega_i$ from the distribution in Equation~\ref{eq:boltzmann-cl}.
The instrumental prescription for this procedure is the inverse coordinate transformation from normal modes back to Cartesian coordinates
\begin{equation}
    \mathbf{R}
    =
    \mathbf{R}_0 + \mathbb{M}^{-\frac{1}{2}} \sum_{i=1}^{N_\mathrm{int}} \Omega_i \mathbf{\Omega}_i,
    \label{eq:sampling}
\end{equation}
where $\mathbb{M}$ represents the diagonal mass matrix.
Thus, by drawing samples of normal coordinates and transforming them, we obtain correctly distributed thermal samples in Cartesian coordinates.

We can now perform thermal NMS by sampling from this auxiliary harmonic ensemble as a source of candidate geometries to be potentially included in the training set of an MLP.
The auxiliary ensemble is thus never used directly and no expectation values are calculated over it.
It only needs to provide good coverage of the thermally accessible region of the PES, which will be the case as long as the harmonic approximation is reasonably accurate for the system of interest.
Specifically, we construct a training set in our active learning procedure by generating a large number of these NMS candidate geometries and screening them in a QbC process using a C-NNP model.
In each QbC iteration, electronic structure calculations are performed only for a small number of selected structures to obtain their potential energies and possibly forces, which then comprise the final training set once the process converges.
The computational cost is thus determined primarily by the geometry optimization procedure, the Hessian calculation, the C-NNP prediction required for screening, and the electronic structure calculations for the selected geometries.
The cost of the sample generation is negligible.
This approach is substantially less computationally demanding when compared to the more conventional approach of sampling the candidate geometries for QbC from an AIMD trajectory, which requires a large number of electronic structure calculations for very similar geometries that do not contribute diversity to the training set.
In contrast to that, NMS generates fully decorrelated geometries by construction, and electronic structure calculations are only needed for the relatively small number of the most important geometries selected by the subsequent QbC process.

So far, we have focused on the situation where NMS is used to sample a classical distribution on the studied PES.
However, since the harmonic approximation describes a molecule as a set of independent one-dimensional harmonic oscillators, we can readily generalize the above classical case to a quantum one as the analytic solution of the quantum harmonic oscillator is known.
Specifically, it is straightforward to show (see Section~\ref{si-sec:theory} of the Supporting Information) that the canonical thermal density of a quantum harmonic oscillator at a given temperature is Gaussian just as its classical counterpart, but broader.
This broadening is encoded in the quantum effective inverse temperature~\cite{Ceriotti2009/10.1103/PhysRevLett.103.030603} 
\begin{equation}
    \beta^*(\beta, \omega)
    =
    \frac{2}{\hbar\omega} \tanh\left( \frac{\beta\hbar\omega}{2} \right)
    \label{eq:betastar}
\end{equation}
at which a classical harmonic oscillator would have the same thermal width as a quantum harmonic oscillator at a reference inverse temperature $\beta$.
Since $\beta^*$ is by definition a frequency-dependent quantity, one cannot describe the whole molecule by a single quantum effective temperature, but instead has to assign one to each individual mode.
In turn, the quantum thermal density is given by
\begin{equation}
    \rho_\mathrm{q}(\Omega_1, \dots, \Omega_{N_\mathrm{int}})
    \propto
    \prod_{i=1}^{N_\mathrm{int}} \exp\left[-\frac{1}{2} \beta^*(\beta, \omega_i) \omega_i^2 \Omega_i^2\right].
    \label{eq:boltzmann-q}
\end{equation}
This simple modification allows one to generate an auxiliary quantum ensemble at practically the same cost as the classical one that would otherwise need to be approached from a significantly more demanding perspective, perhaps based on sampling techniques using the imaginary time path integral formalism.

\subsection{Transition tube sampling}

\begin{figure}[tb!]
    \centering
    \includegraphics{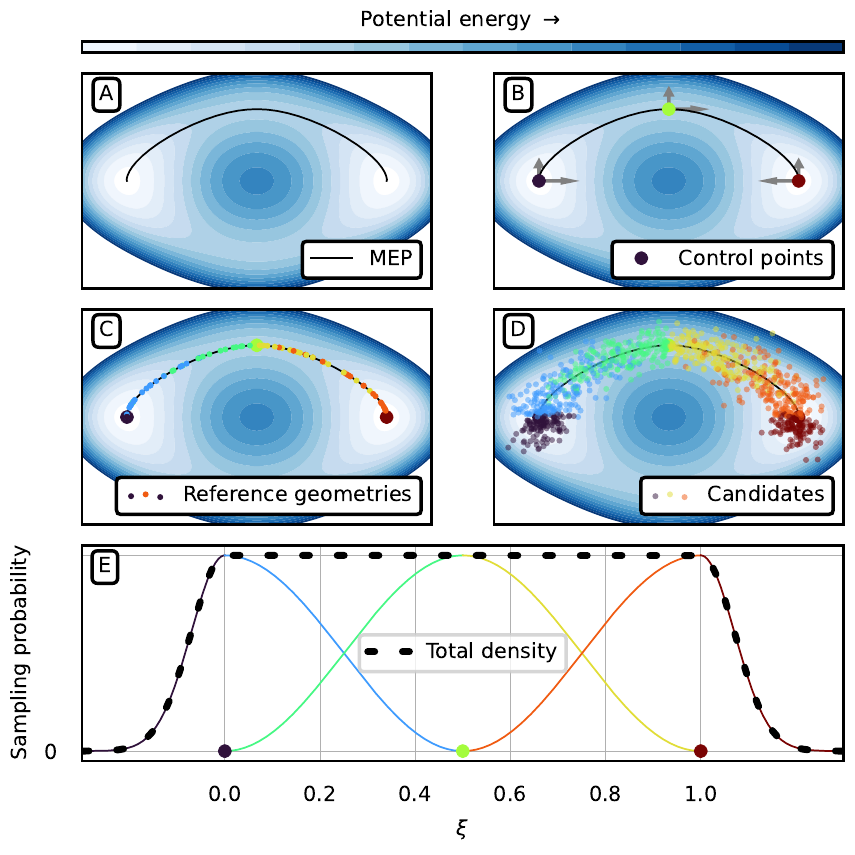}
    \caption{A schematic depiction of the TTS approach proposed for reactive systems.
    Panel A: An illustrative MEP winding through a model two-dimensional configuration space.
    Panel B: Control points are selected and their local normal modes (gray arrows) are calculated. Here, the control points are taken as the two endpoint minima on the MEP (purple and brown dots) and the transition state (yellow--green).
    Panel C: A much denser set of uniformly distributed reference geometries is generated along the MEP.
    Panel D: Each of the reference geometries is distorted using the local modes of their assigned control point (as detailed in panel E) to become a candidate geometry.
    Panel E:
    A detailed view showing how each reference geometry is assigned to a set of local modes.
    A set of reference geometries on the MEP is assigned to each control point following Equation~\ref{eq:probability}.
    At the MEP edges, standard Gaussian thermal NMS is performed outside of the reaction coordinate (decaying purple and brown tails of the total density).}
    \label{fig:wurst}
\end{figure}

Up to this point, we have relied on the ability of the harmonic expansion around a single minimum to approximate the real PES so that the generated samples cover sufficiently all the relevant regions for the purpose of generating an MLP.
Arguably, this is a reasonable requirement for most stable molecules with a single minimum geometry where the onset of the anharmonic region connected to the dissociation of the molecule is not thermally accessible.
On the other hand, it is a stringent requirement for molecular systems which display conformational changes or chemical reactivity and, therefore, are represented by multiple PES minima connected by MEPs: features not captured by a single harmonic expansion.
However, in such cases, it is desirable for the resulting model to be able to describe the potential energy landscape not only around local potential energy minima, but also in the transition regions.
This is vital in the case of low-$k_\mathrm{B}T$ barriers, where spontaneous transitions occur during direct MD.
Nonetheless, it cannot be omitted even in the case of high-$k_\mathrm{B}T$ barriers where an enhanced-sampling simulation would be required to cross the barrier.
Even if the transition does not actually occur, the presence of the transition state may introduce substantial anharmonicity within the original PES basin that an eventual MLP should learn.
However, in the case of barrier transitions it is not desirable to attempt to build the C-NNP model starting from a candidate set representing the true thermal ensemble, even if we could obtain it, since this would lead to a possibly detrimental under-representation of the high-energy configurations close to the transition state in the resulting candidate set and, in turn, to poor performance of the resulting model in the transition regions.

Therefore, we propose the TTS method: a generalization of NMS for systems with transitions that employs local normal modes along an MEP to sample uniformly along the path and with proper thermal weights in all perpendicular directions.
This leads to an auxiliary harmonic ensemble that differs significantly from the true thermal one but enables construction of MLPs with uniform accuracy along the whole transition.
The TTS method naturally reduces to thermal NMS as described above for single-minimum systems in the zero MEP length limit.
The process, illustrated in Figure~\ref{fig:wurst}, starts by finding the MEP $\mathbf{R}(\xi)$ on the given PES (panel A).
Here, $\xi$ is a dimensionless reaction coordinate along the MEP curve through configuration space normalized to the interval from 0 to 1.
In the following, we shall assume that the MEP is available as a continuous, differentiable function of the parameter $\xi$.
In practice, this can be achieved by spline fitting of the discretized representation of the MEP originating from, for instance, a nudged elastic band calculation.~\cite{Jonsson1998/10.1142/9789812839664_0016}
Note that by definition, the MEP is a minimum of the PES in all directions perpendicular to it.
Once the relevant MEP is known, we continue by selecting a sparse set of control points $\mathbf{R}_c, c = 1, \dots, N_\mathrm{p}$ along the MEP at positions $\xi_c$ for which the Hessian matrices are calculated and diagonalized to give the set of local normal mode vectors $\mathbf{\Omega}_{c, i}$ and their corresponding frequencies.
For instance, this can be the two end-point minima and the transition state between them (Figure~\ref{fig:wurst}, panel B), although there is no constraint on how densely one might select the control points along the MEP other than the limiting computational expense of the Hessian matrix calculation.
Formally, the expansion of the PES along the MEP becomes exact under the harmonic approximation in the limit of a large number of control points $N_\mathrm{p}$.
Since we want to achieve uniform sampling along the MEP, we now proceed to the generation of reference geometries on the MEP that do have this property.
Specifically, to each control point $\mathbf{R}_c$ we first assign a probability distribution $p_c(\xi)$ defined as
\begin{equation}
    p_c(\xi)
    =
    \begin{cases}
        \sin^2\left[ \frac{\pi}{2|\xi_c - \xi_{c-1}|} 
        (\xi - \xi_{c-1}) 
        \right], 
        \xi_{c-1} \leq \xi \leq \xi_c \\
        \cos^2\left[ \frac{\pi}{2|\xi_c - \xi_{c+1}|} 
        (\xi - \xi_{c}) 
        \right], 
        \xi_{c} \leq \xi \leq \xi_{c+1} \\
        0 \ \text{otherwise}
    \end{cases}
    \label{eq:probability}
\end{equation}
for all $\xi \in [0, 1]$ so that the identity
\begin{equation}
    p(\xi)
    =
    \sum_{c=1}^{N_\mathrm{p}}
    p_c(\xi)
    =
    1
\end{equation}
holds over the whole length of the MEP (Figure~\ref{fig:wurst}, panel E over the range of $\xi$).
Note that the choice of squares of harmonic functions is only one out of many possibilities, and any other pair of complementary functions that sum up to unity would work in this case.
Next, we generate an arbitrary number of reference geometries $\mathbf{R}_0(\xi)$ at a chosen linear density by drawing random values of $\xi$ from the above distributions and passing them to the continuous prescription of the MEP, all while keeping track of the parent $c$-th control point (Figure~\ref{fig:wurst}, panel C).
Analogously to the distortion of the minimum geometry in the single-minimum case through Equation~\ref{eq:sampling}, we distort each of these reference geometries using the normal modes and frequencies of its parent control point using
\begin{equation}
    \mathbf{R}
    =
    \mathbf{R}_0(\xi)
    +
    \mathbb{M}^{-\frac{1}{2}}
    \sum_i{}^{'}
    \Omega_{c, i}
    \left[
        1 - \mathbb{P}(\xi)
    \right]
    \mathbf{\Omega}_{c, i},
\end{equation}
where the normal coordinate values $\Omega_{c, i}$ are sampled thermally according to Equation~\ref{eq:boltzmann-cl} or \ref{eq:boltzmann-q} (Figure~\ref{fig:wurst}, panel D); the prime indicates that modes with imaginary frequencies are omitted from the sum.
The matrix $\mathbb{P}(\xi)$ is the projector on the tangent direction at point $\xi$ which can be constructed analytically from $\dd \mathbf{R}(\xi) / \dd \xi$.
This is used to obtain distortions strictly perpendicular to the MEP and thus to correct for the approximate validity of the normal mode expansion calculated at $\xi_c$ for all the displaced geometries.
However, the use of the decaying probability distributions (Equation~\ref{eq:probability}) favors the use of the local modes close to their origin.
Through this procedure, one obtains a set of candidate geometries distributed inside a tube around the MEP the width of which is given thermally.
At this point, this tube still has open ends cut sharply by the planes defined by normal vectors equal to the MEP tangent vector at the endpoints of the MEP.
Since these points are (usually) also well-defined minima on the PES, the presence of these sharp edges is easily sanitized by appending the usual thermal NMS samples at these minima, although only adding the configurations away from the MEP (Figure~\ref{fig:wurst}, decaying gray dashed lines).
In other words, just one half of the multivariate Gaussian is appended to the tube that does not overlap with it.
In our TTS implementation, we ensure that the uniform density of the sampling along the MEP and the one at the peak of the half-Gaussian are seamlessly matched (as described in Section~\ref{si-sec:theory} of the Supporting Information).

Using the described sampling approach leads to an auxiliary ensemble of candidates that does not correspond to the true thermal ensemble, but contains a balanced selection of geometries distributed uniformly along the MEP with classical or quantum thermal displacements around it.
Just like with plain NMS, we submit these samples as candidates to the QbC procedure, where in each iteration a large number of them is screened and a small number of those is selected to be included in the training set.
\textit{Ab initio} calculations are only required for these selected geometries.
This enables the building of diverse training sets in which all representative structures that might be encountered in a future simulation are contained so that the resulting model is, in fact, able to accurately describe the PES along the whole MEP, even in regions that have negligible thermal populations.
Similar to NMS, the computational cost of TTS is determined primarily by the MEP optimization procedure, the Hessian calculation, the C-NNP prediction required for screening, and the electronic structure calculations for the selected geometries.
In general, this can be expected to be substantially less computationally demanding than executing direct, or even enhanced-sampling, classical or path integral AIMD simulations and sampling from their trajectories.

\section{Computational details}
\label{sec:computational-details}

\subsection{\textit{Ab initio} electronic structure}

Two different levels of electronic structure theory were used in the simulations presented in this work.
In both cases, we used the implementation provided by the CP2K software package~\cite{Hutter2014/10.1002/wcms.1159} with its Quickstep DFT module~\cite{Vandevondele2005/10.1016/j.cpc.2004.12.014,Kuehne2020/10.1063/5.0007045}.
We described the electronic structure of the benzene molecule in the gas phase at the self-consistent charge density-functional tight binding~\cite{Porezag1995/10.1103/PhysRevB.51.12947} (SCC-DFTB) level with third-order diagonal corrections.
The system was enclosed in a \SI{10}{\angstrom} wide cubic box with open boundary conditions.
For malonaldehyde and DABQDI systems, we used the revPBE0-D3 hybrid density functional~\cite{Perdew1996/10.1103/PhysRevLett.77.3865,Zhang1998/10.1103/PhysRevLett.80.890,Adamo1999/10.1063/1.478522,Goerigk2011/10.1039/c0cp02984j} combined with the TVZ2P Gaussian basis set~\cite{Vandevondele2005/10.1016/j.cpc.2004.12.014,Guidon2008/10.1063/1.2931945,Guidon2009/10.1021/ct900494g} to represent the molecular orbitals and a plane wave basis with a \SI{600}{Ry} cutoff to represent the density.
The core electrons of the heavy atoms were represented using Goedecker--Tetter--Hutter pseudopotentials~\cite{Goedecker1996/10.1103/PhysRevB.54.1703}.
In addition, we used the auxiliary density matrix method~\cite{Guidon2010/10.1021/ct1002225} with the cpFIT3 fitting basis set for the DABQDI molecule.
Both systems using hybrid DFT were centered in a \SI{15}{\angstrom} wide cubic box with open boundary conditions and the wavelet Poisson equation solver.

\subsection{C-NNP model generation}

Throughout all of our investigations, we used committees consisting of eight different Behler--Parrinello neural network potentials~\cite{Behler2007/10.1103/PhysRevLett.98.146401}, where for each of them, a different initialization of weights and a different 90 \% subset of the full training data set was used to ensure a diverse committee.
The models consisted of two hidden layers of 20 nodes each and were trained using the multistream~\cite{Singraber2019/10.1021/acs.jctc.8b01092} adaptive extended Kalman~\cite{Shah1992/10.1016/S0893-6080(05)80139-X,Blank1994/10.1002/cem.1180080605} filter with 32 streams.
The input features were a standard set of atom-centered symmetry functions~\cite{Schran2021/10.1073/PNAS.2110077118}.
The training of the model was done using the n2p2 package~\cite{Singraber2019/10.1021/acs.jctc.8b01092} and the selection of training structures by QbC was done with a development version of our AML package~\cite{AMLgithub}.
For benzene, 20 structures were randomly sampled initially and in each of the 40 QbC iterations, 10 new structures with the highest committee force disagreement were added to the data set, for a total of 420 training geometries.
The final NNPs were trained for 2000 epochs.
For the first generation of malonaldehyde and DABQDI models, 20 initial structures were sampled randomly and then the 15 structures with the highest disagreement were selected in each of the 40 QbC iterations, for a total of 620 training geometries.
For malonaldehyde, where additional generations of models were required (as detailed in Section~\ref{sec:results}), the original training set was supplemented by additional structures QbC-sampled from an MD trajectory which was produced using the previous C-NNP model.
Here, 15 structures were added in each iteration until the force committee disagreements of the selected structures and the remaining candidate structures were similar.
All reference calculations were done using CP2K and the electronic structure settings described above.

\subsection{Geometry optimization and vibrational analysis}

The optimization of the minimum reference geometries for benzene and DABQDI was executed natively in the CP2K software.
It was performed using the BFGS optimizer~\cite{Broyden1970/10.1093/imamat/6.1.76} combined with threshold criteria of \SI{0.07}{eV \ \angstrom^{-1}} for the maximum change in force components, \SI{0.009}{\angstrom} for the change in atomic positions and \SI{0.13}{eV} for the change in total energy.
For the malonaldehyde molecule, we employed the Atomic Simulation Environment (ASE)~\cite{Larsen20177/10.1088/1361-648X/aa680e} together with CP2K and performed the optimization using the FIRE optimizer~\cite{Bitzek2006/10.1103/PhysRevLett.97.170201} while specifying only a force criterion of \SI{0.01}{eV \ \angstrom^{-1}}.
Additional constrained optimizations in the case of DABQDI needed for the relaxed PES scan were performed using the constraint functionality provided by ASE together with its FIRE optimizer.
The Hessian matrix evaluation on the optimized structures was performed in each case using CP2K and a Cartesian atomic displacement of \SI{0.0005}{\angstrom}.

\subsection{Nudged elastic band calculations}

The relevant MEPs needed for the TTS procedure were obtained through the climbing-image~\cite{Henkelmann2000/10.1063/1.1329672} nudged elastic band~\cite{Jonsson1998/10.1142/9789812839664_0016} (CI-NEB) optimization procedure as implemented in CP2K.
The initial band geometries in this work consisted of 15 replicas of the molecule in question including the two fixed, pre-optimized endpoints, and were obtained through linear interpolation.
The spring constant of the harmonic links between the neighboring replicas was kept constant at the value of \SI{4.86}{eV \ \angstrom^{-2}}.
We used a force convergence criterion of \SI{0.007}{eV \ \angstrom^{-1}} and the minimization of the band energy was performed using a DIIS optimizer.

\subsection{MD simulations}

All MD simulations involving both \textit{ab initio} as well as C-NNP potentials~\cite{Schran2020/10.1021/acs.jctc.9b00805} were run using the CP2K package.
The simulations with classical representation of the nuclei were propagated at a temperature of \SI{300}{K} using a time step of \SI{0.5}{fs} to numerically integrate the Langevin equation with the friction coefficient $\gamma$ of \SI{0.02}{fs^{-1}} to achieve canonical sampling. 
The path integral simulations that include nuclear quantum effects were performed using imaginary-time ring polymers consisting of 64 replicas using the RPMD propagator.
The canonical distribution at \SI{300}{K} was sampled using the path integral Langevin equation thermostat~\cite{Ceriotti2010/10.1063/1.3489925} with the time constant for the centroid motion of \SI{200}{fs} while the integration time step was kept at \SI{0.25}{fs}.

\subsection{Umbrella sampling}

The initial conditions for each umbrella sampling window were extracted from a steered MD trajectory, which was performed in the CP2K v2022.1 software package combined with the PLUMED plugin~\cite{Bonomi2009/10.1016/j.cpc.2009.05.011,Bonomi2019/10.1038/s41592-019-0506-8,Tribello2014/10.1016/j.cpc.2013.09.018}.
In this case, the value of the proton-sharing coordinate $\delta_1$ (as detailed in Section~\ref{sec:results}) was biased from \SI{-1.2}{\angstrom} to \SI{1.2}{\angstrom} during a \SI{10}{ps} long simulation using a moving harmonic restraint with the force constant $\kappa$ of \SI{500.0}{kJ \ mol^{-1} \ \angstrom^{-2}}.
The simulation was performed classically with an integration time step of \SI{0.5}{fs} in the canonical ensemble at \SI{300}{K} using a local CSVR thermostat~\cite{Bussi2007/10.1063/1.2408420} with a time constant of \SI{50}{fs}.

30 equidistant umbrella sampling windows separated by \SI{0.08}{\angstrom} were set up from the above steered MD simulation.
Individually in each window, the value of $\delta_1$ was biased by a static harmonic restraint of \SI{500.0}{kJ \ mol^{-1} \ \angstrom^{-2}} and simulated for \SI{50}{ps} using the same setup as for the steered MD simulation above.
The overlap of the corresponding histograms of $\delta_1$ values observed in each simulation window is shown in Section~\ref{si-sec:additional-results} of the Supporting Information.
The value of $\delta_2$ was kept unbiased in each simulation window but was monitored for use in the following analysis.
The biased configurations were reweighed to the unbiased ensemble using a Python implementation of the multistate Bennet acceptance ratio~\cite{Shirts2008,WHAMgithub} (MBAR) procedure to obtain both a one-dimensional free energy profile for the proton-transfer along $\delta_1$ as well as a two-dimensional free energy surface showing the dependence on both proton-sharing coordinates.
This was done by determining the thermal weight associated with each configuration in the biased simulations and using these to obtain the probability distribution in the $\delta_1$, $\delta_2$ subspace, and from that the corresponding free energy surface.

\section{Results and discussion}
\label{sec:results}

To showcase the performance of the TTS procedure in the creation of models for realistic potentials, we select three different gas-phase molecules with increasing complexity of their PES.
We begin with benzene, which represents a single-minimum system with a close-to-harmonic potential at room temperature and thus allows us to illustrate the simple thermal NMS procedure. 
This is followed by a study of the enol form of 1,3-propanedial (malonaldehyde), which exhibits reactivity by sharing the acidic proton between the two oxygen atoms spontaneously at ambient conditions~\cite{Tuckerman2001/10.1103/PhysRevLett.86.4946}.
Finally, we focus on a more involved proton-sharing system represented by 2,5-diaminobenzoquinone-1,4-diimine (DABQDI), which has two proton-sharing sites~\cite{Cahlik2021/10.1021/acsnano.1c02572}.
Spontaneous proton transfer is hindered by a barrier thermally insurmountable at room temperature, and an enhanced sampling simulation is necessary to determine the free energy profile.

\subsection{Benzene}

\begin{figure}[tb!]
    \centering
    \includegraphics{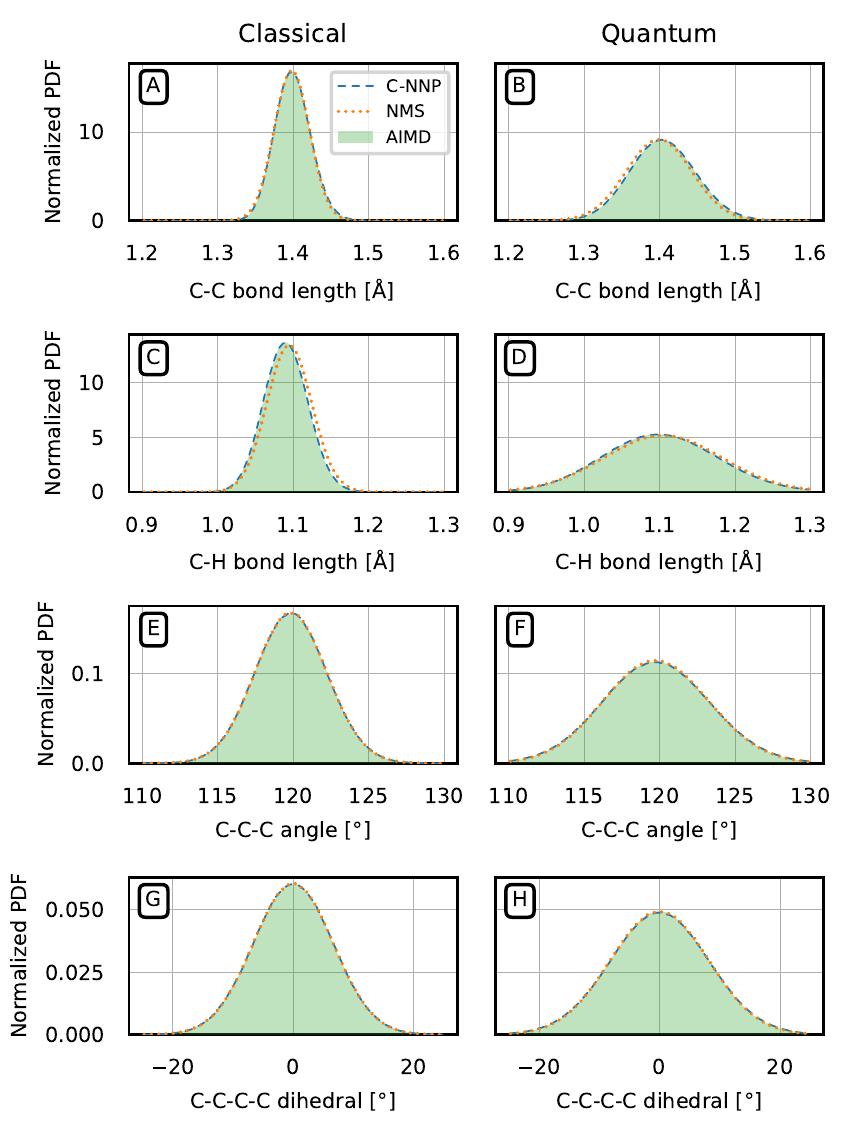}
    \caption{
    Thermal geometry properties of benzene in the gas phase at \SI{300}{K} from classical MD (left column) and path integral MD (right column) compared between simulations using the reference DFTB potential, the harmonic TTS ensemble, and simulations using a C-NNP model building on the thermal NMS geometries.
    Panels A and B show the distribution of C--C bond lengths, panels C and D the distribution of C--H bond lengths, panels E and F the distribution of C--C--C angles, and, finally, panels G and H the distribution of the C--C--C--C dihedral angles.
    }
    \label{fig:benzene}
\end{figure}

To lead off the discussion of the ability of TTS to seed a training set for the creation of C-NNP models for realistic systems in the gas phase, we focus on the benzene molecule.
It represents an ideal example to illustrate the basic idea of thermal NMS using a single normal mode expansion at an optimal geometry, since it features a single configurational minimum and the surrounding PES exhibits almost no anharmonic effects.

To prepare the ground for comparison with the relevant C-NNP data, we initially performed one \SI{250}{ps} AIMD simulation of gas-phase benzene at \SI{300}{K} at the DFTB level using a classical representation of the atomic nuclei as well as a \SI{100}{ps} PIMD simulation using 64 replicas to approximate the imaginary time path.
Two C-NNP models were then based on candidate sets obtained from a thermal NMS of gas-phase benzene using a Hessian matrix calculated at the same DFTB level of theory as the (PI)-AIMD simulations at \SI{300}{K} for the classical model and with the appropriate effective temperatures at \SI{300}{K} for the quantum one.
The resulting models were evaluated on test sets consisting of 1000 structures sampled from the two AIMD trajectories.
Both models performed very well with an energy root mean square error (RMSE) of \SI{1.66}{meV} and \SI{5.90}{meV} for the model constructed for the use without and with path integral structures, respectively.
The RMSE for a single force component was \SI{14.9}{meV \angstrom ^ {-1}} and \SI{30.4}{meV \angstrom ^ {-1}}. 
Subsequently, the models were used to obtain new \SI{500}{ps} long MD and \SI{100}{ps} PIMD simulations at \SI{300}{K}.

The comparison of the C-NNP models to the corresponding (PI)-AIMD trajectories in terms of molecular geometry properties is summarized in Figure~\ref{fig:benzene}.
In general, we can see the expected broadening of probability distributions due to nuclear quantum effects when we compare the left and right columns of Figure~\ref{fig:benzene}.
In both the classical and quantum case, we observe a perfect match between the \textit{ab initio} (green shading) and C-NNP distributions (blue dashed lines) in C--C bond lengths (panels A and B), C--H bond lengths (panels D and E), C--C--C angles (panels E and F), and C--C--C--C dihedrals (panels G and H).
The two types of covalent bonds have expected distributions; the mean of the C--C--C angle is located at 120$^\circ$ which shows the average hexagonal arrangement of the aromatic ring subject to planarity, which is, in turn, demonstrated by the (signed) C--C--C--C dihedral angle peaking at 0$^\circ$ as expected.
This level of agreement suggests that the final models used for production MD represent excellent approximations to the original DFTB PES.
Additionally, we show the distributions of the NMS structures (orange dotted lines) alongside the anharmonic distributions.
These exhibit significant overlap with both the (PI)-AIMD and C-NNP data.
This suggests that the harmonic approximation to the original ensemble is relatively good and confirms the assumed high degree of harmonicity of the \SI{300}{K} gas-phase benzene PES, even in the quantum case.
However, note that the match of the NMS data with the (PI)-AIMD data is not nearly as perfect as that of the C-NNP data and certain deviations are, in fact, present.
As discussed in Section~\ref{sec:theory}, these are to be expected since the NMS ensemble is only auxiliary and its goal is to provide sufficient coverage of the accessible PES which ultimately leads to an accurate C-NNP model.
Using thermal NMS, we were thus able to construct a C-NNP that accurately describes the original PES of benzene based on a single Hessian evaluation and 420 single-point electronic structure calculations.

\subsection{Malonaldehyde}

The enol form of malonaldehyde is a simple organic molecule that has been used in MD simulations~\cite{Tuckerman2001/10.1103/PhysRevLett.86.4946} to illustrate a simple intramolecular proton-transfer reaction
\begin{equation*}
\schemestart
    \setchemfig{atom sep=1.5em}
    \chemfig{(=[:90]O)-[:-30]=[:30](-[:90]O(-[:150]H))}
    \arrow{<=>}
    \setchemfig{atom sep=1.5em}
    \chemfig{(-[:90]O(-[:30]H))=[:-30]-[:30](=[:90]O)}
\schemestop
\label{eq:malonaldehyde}
\end{equation*}
in which the proton is moved from the enol oxygen to the aldehyde oxygen with a simultaneous electronic rearrangement which altogether causes the reactant and the product to become symmetrically mirrored, chemically identical structures.
As such, malonaldehyde is a convenient molecule to demonstrate the ability of TTS to describe a simple reaction.

\begin{figure}[tb!]
    \centering
    \includegraphics{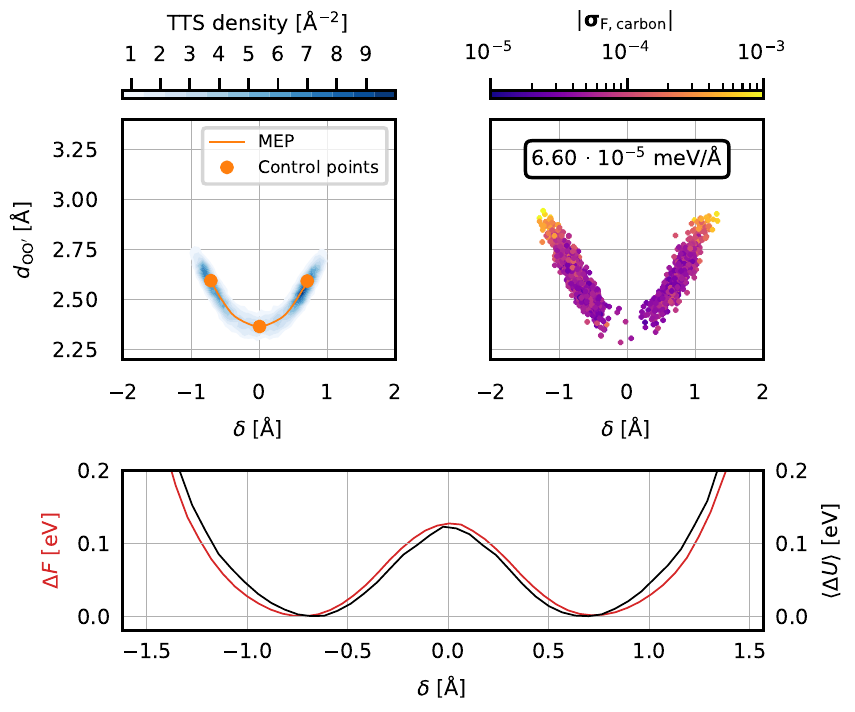}
    \caption{
        The TTS sampling of the malonaldehyde proton transfer and the MD simulation with the resulting C-NNP model.
        The top left panel shows the relevant MEP (orange) with the three selected control points in the two configurational minima and the transition state highlighted and the distribution of the TTS geometries (blue).
        The top right panel shows a scatter plot of a subset of geometries (selected with a stride of \SI{37.5}{fs}) obtained during a \SI{250}{ps} MD simulation using the C-NNP model built on top of the TTS ensemble.
        Each point is colored by the norm of the force committee disagreement on the carbon atoms and the mean of the quantity is shown in the box.
        Note the high force disagreement in the high $\delta$ tails of the distribution.
        The bottom panel shows the Boltzmann-inverted free energy profile (red) and the corresponding binned average potential energy of the system (black, aligned to zero) along the proton transfer reaction as observed in the MD simulation.
    }
    \label{fig:malonaldehyde1}
\end{figure}

With the aim to describe the proton-sharing process accurately, we decided to model the original \textit{ab initio} PES at the hybrid DFT level using the revPBE0-D3 functional.
Thanks to the ability to use quantum normal coordinate distributions in the TTS method, we produced both classical and quantum models and classical and PIMD trajectories  for malonaldehyde to test and showcase this functionality.
However, we focus mostly on the classical case in the main text and discuss the complementary quantum results, for which qualitatively similar conclusions arise, in the Supporting information, Section \ref{si-sec:pimd-for-malonaldehyde}.
The starting point of the TTS procedure is the proton-sharing MEP, which was discretized into 15 replicas and optimized using the CI-NEB procedure and the revPBE0-D3 density functional.
For illustration purposes, we decided to use the full MEP with both the reactant and product explicitly represented: this is strictly speaking not necessary since the symmetry of the reaction allows the use of only one non-trivial half of the MEP for TTS.
Out of the optimized full-length MEP, three control points were selected in the two minima (reactant and product) and in the transition state.
For the visualization of the multidimensional configuration data, we choose the reduction into a 2D space of two geometric parameters: the proton-sharing coordinate $\delta(\mathbf{R}) = |\mathbf{R}_{\mathrm{O}} - \mathbf{R}_{\mathrm{H}}| - |\mathbf{R}_{\mathrm{O'}} - \mathbf{R}_{\mathrm{H}}|$ and the oxygen--oxygen $d_{\mathrm{OO'}}(\mathbf{R}) = |\mathbf{R}_{\mathrm{O}} - \mathbf{R}_{\mathrm{O'}}|$ where O and O' denote the two oxygen atoms that share the proton H.
The parameters are illustrated in the snapshot on the left of Figure~\ref{fig:malonaldehyde1}.
The obtained optimized MEP and the selected control points in this representation are shown in the top left panel of Figure~\ref{fig:malonaldehyde1} in orange.
The TTS classical candidate structures were then generated using the procedure outlined in Section~\ref{sec:theory} at the temperature of \SI{300}{K}, linear sampling density along the MEP of \SI{1e5}{\angstrom ^ {-1}} and matched-density sampling at the minima.
The distribution of the obtained configurations is shown in the top left panel of Figure~\ref{fig:malonaldehyde1} as blue contours.
The same distribution colored by the assignment of each candidate to the individual control points (corresponding to the situation shown in panel D of Figure~\ref{fig:wurst}) is shown in Section~\ref{si-sec:additional-results} of the Supporting information.
Note that this particular presentation of the data does not do justice to the uniformity of the sample distribution along the MEP as the regions around the minima seem to be more populated than the transition state.
This is an effect of the deformation of the configuration space by the projection on the selected subspace; the samples are distributed uniformly in the full dimension.
After passing the resulting set of candidates through the QbC selection and training a C-NNP model on the obtained training set, the model was used to run a direct \SI{250}{ps} MD simulation of gas-phase malonaldehyde at \SI{300}{K}.
A subset of the obtained configurations is shown in the form of a scatter plot in the top right panel of Figure~\ref{fig:malonaldehyde1} colored by the decadic logarithm of the norm of the force committee disagreement on carbon atoms in the usual $\delta$ and $d_{\mathrm{OO'}}$ representation.
Additionally, a 1D free-energy profile obtained by a Boltzmann inversion of the probability density of configurations along the $\delta$-axis is shown in the bottom panel of Figure~\ref{fig:malonaldehyde1}; the height of the barrier is approximately \SI{120}{meV} which corresponds to roughly 5~$k_\mathrm{B}T$ at \SI{300}{K}.
This accounts for the expected low, but existing population surrounding the transition state at $\delta=$ \SI{0}{\angstrom}. Alongside the free energy profile, we show the corresponding average potential energy as a function of $\delta$.

\begin{figure*}[tb!]
    \centering
    \includegraphics{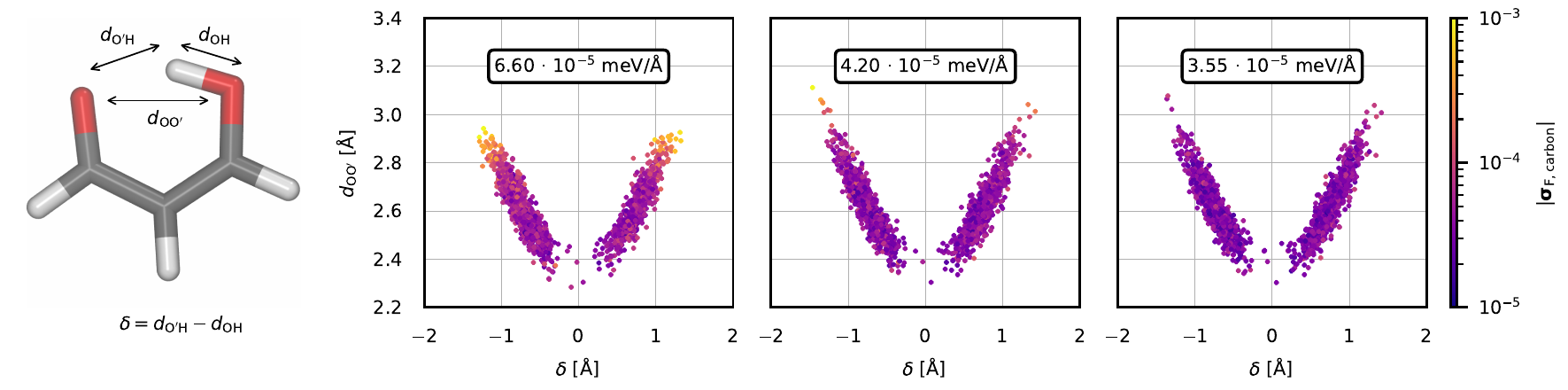}
    \caption{
        Evolution of the force disagreement of the carbon atoms through multiple instances of QbC.
        The left panel shows a subset of configurations originating from an MD simulation using a model trained on data selected directly from the TTS candidate set (identical data as in the top right panel of Figure~\ref{fig:malonaldehyde1} is shown).
        The force disagreement (depicted using the color scale) in the vicinity of the two configurational minima and along the proton-sharing reaction is adequately low, however, for structures with a high absolute $\delta$ it is more than one order of magnitude higher.
        The central panel shows configurations and disagreements obtained from an MD simulation using a model trained on the initial training data augmented by QbC-selected high-disagreement configurations from the data in the left panel.
        In turn, the right panel shows data obtained by improving the model by the new data sampled in the simulation shown in the central panel.
        Most notably, structures at the tails of the populated configuration space are substantially improved.
        The mean force disagreement over all configurations in each data set is shown in the framed box in each panel.
        The $\delta$ and $d_{\mathrm{OO'}}$ coordinates are illustrated in the snapshot to the left of the panels.
    }
    \label{fig:malonaldehyde2}
\end{figure*}

An important observation can be made from the presented data.
By comparison of the two distributions in the top panels of Figure~\ref{fig:malonaldehyde1}, it is clear that the TTS distribution populates a smaller volume of the configuration space than the data obtained from the MD simulation.
In this particular case, it means that TTS does not directly provide good enough coverage and the resulting model is undertrained in the missing, yet thermally accessible regions.
Specifically, the C-NNP model performs poorly in the large $d_{\mathrm{OO'}}$ tails of the shown distribution, as quantified by the larger force disagreement values.
On the contrary, in regions around the proton-sharing MEP, the coverage is good and the force disagreement remains small, despite the tiny thermal population.
Regardless of the elevated model uncertainty for some configurations, these MD simulations remain stable. 
The observed increased disagreement can be interpreted in the following way: going in the opposite direction from the minima as the proton-sharing MEP, the true anharmonic PES has a potential wall that grows slower than the harmonic wall captured by the TTS distribution and, therefore, the thermal coverage of the TTS configurations cannot reach far enough.
In principle, this behavior is caused by either the strongly anharmonic character of the chemical bonds leading to bond dissociation or, more likely in this case, the presence of another reactive process leading to a new transition state.
In the following two paragraphs, we present two possible solutions to the issue.
The first one relies on an active-learning-based iterative improvement of the model, which has the advantage of requiring no knowledge of the origin of the anharmonicity, but is tedious to perform since several intermediate simulations and model generations need to be created.
Meanwhile, the other solution relies on the chemical intuition of the user to identify the reactive nature of the issue with the aim to extend the initial TTS, which leads to a fully capable C-NNP model straight away.

The QbC process can be used to fill in an already existing training data set that has gaps, perhaps due to an incomplete TTS candidate set in the QbC selection for the initial model.
We illustrate this process in Figure~\ref{fig:malonaldehyde2}, where the left panel shows the same data as the top right panel of Figure~\ref{fig:malonaldehyde1} as a starting point.
Regions of configuration space not covered well in the training set of this generation 1 model can be easily identified by the high committee disagreement, as can be seen in the tails of the distribution.
Hence, one can start a new QbC using the existing training data set augmented by structures from an MD simulation performed with the initial C-NNP. 
Depending on the size of the gaps in the initial training data set, only a few iterations of QbC are typically necessary. 
However, adding these structures to the training data set can lead to substantial changes in the previously inaccurate regions of the potential energy surface, resulting in an MD simulation that again reaches new regions of the configuration space where the shape of the potential energy surface is yet unknown to the model and the committee disagreement is high.
This can be seen in the generation 2 model shown in the middle panel of Figure \ref{fig:malonaldehyde2}.
Therefore, multiple repetitions of the MD--QbC cycle might be necessary until a highly accurate model that exhibits low and uniform disagreements over the sampled data is reached, as is the case for the generation 3 model in the right panel of Figure~\ref{fig:malonaldehyde2}.
As such, the approach could become practically cumbersome when the regions of high disagreement coincide with regions of high free energy and long MD simulations are needed to uncover these structures, but nonetheless represents a general solution to the anharmonicity problem.
Overall, repeating the cycle of sampling MD configurations with a given generation of a C-NNP model followed by training a new generation on a training set enhanced by high-disagreement QbC-selected structures from the previous MD simulations leads to a force disagreement that is lower in the problematic PES regions and, therefore, more uniform overall.
In addition, we observe a decrease of the mean of the force disagreement of the sampled configurations due to the fact that the size of the training set increases in each generation.
Specifically, 620 structures were used for the original model in the left panel of Figure~\ref{fig:malonaldehyde2}, 800 structures for the model in the middle panel, and 950 structures for the last model in the right panel.
This approach can be beneficial in systems where it is difficult to identify the origin of the anharmonicity of the original PES, but is rather demanding from the point of view of both computational requirements and user involvement due to the need for the semi-supervised iterative procedure.

\begin{figure}[tb!]
    \centering
    \includegraphics{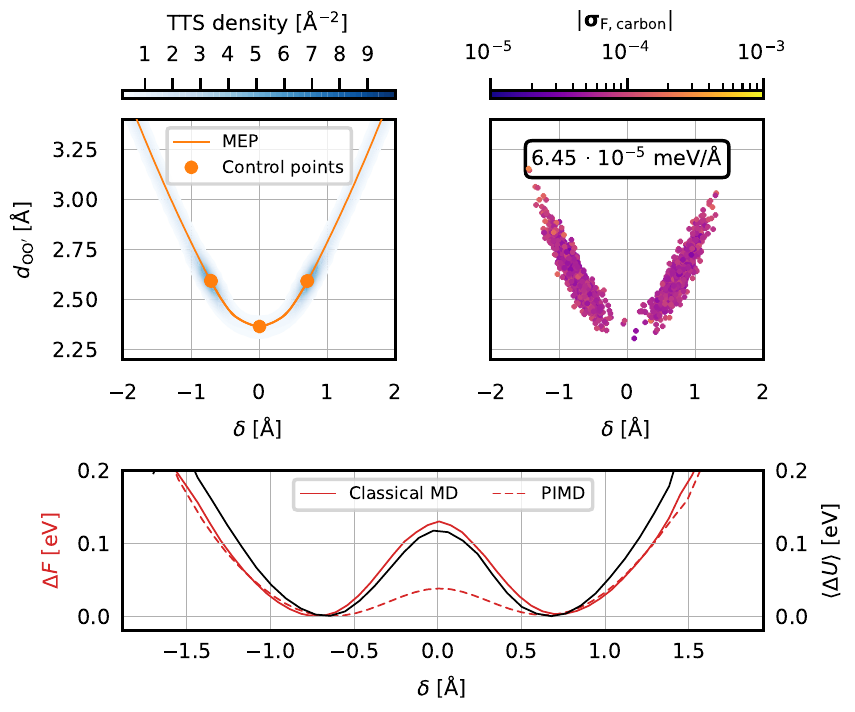}
    \caption{
        TTS sampling of the extended malonaldehyde MEP containing the proton transfer reaction as well as the single C--C bond torsion and the MD simulation with the resulting C-NNP model.
        Identical quantities as in Figure~\ref{fig:malonaldehyde1} are shown with the free energy being plotted here for both classical (solid line) and path-integral (dashed line) data.
        The orange curve shown in the top left panel is a unison of the MEPs corresponding to the proton transfer and the C--C bond torsion; the projection of the latter into the $\delta$, $d_{\mathrm{OO'}}$ subspace is not, strictly speaking, a physically meaningful concept, but clearly visualizes the fact that MEP is the continuation in the desired direction.
        As such, the C-NNP model trained on the combined TTS structures has no further deficiencies as shown by the overall uniform force disagreement in the top right panel.
        }
    \label{fig:malonaldehyde3}
\end{figure}

However, in the case of malonaldehyde, the general active learning iterative procedure to improve the model might be excessive.
The possible reasons for the softer-than-harmonic wall due to conformational flexibility are few and the particular direction against which the first generation of the model is pushing can be easily identified with the \textit{s-cis} and \textit{s-trans} torsional isomerism
\begin{equation*}
\schemestart
    \setchemfig{atom sep=1.5em}
    \chemfig{(=[:90]O)-[:-30]=[:30](-[:90]O(-[:150]H))}
    \arrow{<=>}
    \setchemfig{atom sep=1.5em}
    \chemfig{O=[:30]-[:-30]=[:30](-[:90]O-[:150]H)}
\schemestop
\label{eq:malonaldehyde2}
\end{equation*}
which is mediated by rotation around the C--C single bond in the propane backbone.
The optimized MEP corresponding to this torsion displays a perfect continuation in the correct direction when projected into the $\delta$ and $d_{\mathrm{OO'}}$ subspace, as shown in the top left panel of Figure~\ref{fig:malonaldehyde3} in orange.
Although this pair of descriptors is not appropriate for the whole torsion MEP, which entails a more complicated motion, it is accurate enough at small deviations from the equilibrium geometry.
To include structures along this MEP into the initial (first generation) proton-sharing TTS, two control points were chosen in the minimum (shared by the two MEPs) and in the new transition state (not shown in Figure~\ref{fig:malonaldehyde3}, as it is around $d_{\mathrm{OO'}} =$ \SI{3.8}{\angstrom}).
We do not need to use the \textit{s-trans} minimum at all, as we are not interested in including the transition itself, only the shape of the PES on the side of the global minimum.
A new TTS was performed between these control points with the same parameters as the initial one and the resulting distribution of the combined sets of configurations is shown in blue in the top left panel of Figure~\ref{fig:malonaldehyde3}.
Running a \SI{250}{ps} long MD simulation with a new C-NNP model trained on the QbC-selected training set from this combined candidate set leads to the distribution shown in the top right panel in Figure~\ref{fig:malonaldehyde3}.
Clearly, the distribution reaches all the expected thermally accessible regions, the force disagreement is evened out across the configurations, and the high-disagreement tails are no longer present.
The mean value of the disagreement is comparable to that of the first-generation model, as these two models are based on training sets of the same size.
The bottom panel of Figure~\ref{fig:malonaldehyde3} shows the potential energy and the free energy profile along $\delta$, where the latter is shown for both classical and path-integral data.
In all three curves, one can observe a softening of the barrier in the high $|\delta|$ regions in comparison to the data shown in Figure~\ref{fig:malonaldehyde1} resulting from the present C-NNP being aware of the anharmonic nature of the PES in these regions. 
The transition-state free energy of \SI{43}{meV} for the quantum simulation is $\sim$4 times lower than in the classical case, which is a manifestation of proton tunneling through the barrier.
This is qualitatively consistent with existing literature but quantitatively different from the results reported for the BLYP functional, where the classical free energy barrier is somewhat higher and the quantum effect is substantially smaller, decreasing the barrier by a factor of only $\sim$2~\cite{Tuckerman2001/10.1103/PhysRevLett.86.4946}.

For further insights, a test set independent of the training set data was created by generating 500 structures using TTS and sampling 500 structures from the classical MD trajectory shown in Figure~\ref{fig:malonaldehyde3}, and evaluating their energies and forces with the revPBE0-D3 reference method.
The generation 3 C-NNP of the iterative approach as well as the extended TTS C-NNP perform well with an energy root mean square error (RMSE) of \SI{1.80}{meV} and \SI{3.44}{meV} and a force component RMSE of \SI{18.3}{meV \angstrom ^ {-1}} and \SI{24.4}{meV \angstrom ^ {-1}}, respectively.
The slightly elevated RMSE of the extended TTS approach is due to the broader coverage of the training set.
It includes a range of geometries along the C--C single bond torsion, even in regions that are not populated during MD, leading to a less dense coverage of the rest of the configuration space.
The validation errors of the intermediate models of the iterative approach and the distribution of errors within configuration space are discussed in more detail in the Supporting Information, Section \ref{si-sec:validation-of-malonaldehyde}.

Both of the approaches above thus yield highly accurate models for the description of the proton-sharing reaction in malonaldehyde for classical and quantum nuclei.
The advantage of one over the other therefore depends mostly on the specific situation in which the need for any of them should arise: if the reaction coordinate of the complementary transition can be identified, then the latter approach using the extended TTS reaches the desired result with higher efficiency.
Note that this approach can also be used when multiple different transitions are to be included in a single model.

\subsection{DABQDI}

The most complex reactive system used to demonstrate the performance of the TTS method is represented by the DABQDI molecule.
This nitrogenated benzoquinone derivative can exchange two protons between the neighboring amine and imine groups
\begin{equation*}
\schemestart
    \setchemfig{atom sep=1.5em}
    \chemfig{HN=[:30]-[:-30]=[:30](-[:-30]NH_2)-[:90](=[:30]NH)-[:150]=[:210](-[:150]H_2N)-[:270]}
    \arrow{<=>}
    \setchemfig{atom sep=1.5em}
    \chemfig{H_2N-[:30]=[:-30]-[:30](=[:-30]NH)-[:90](-[:30]NH_2)=[:150]-[:210](=[:150]NH)-[:270]}
\schemestop
\label{eq:qabqdi}
\end{equation*}
again accompanied by an electronic rearrangement that maintains the $\pi$-electron conjugation throughout the process.
However, this time, the proton-sharing does not take place at ambient conditions, which suggests high barriers for the process.

\begin{figure*}[tb!]
    \centering
    \includegraphics{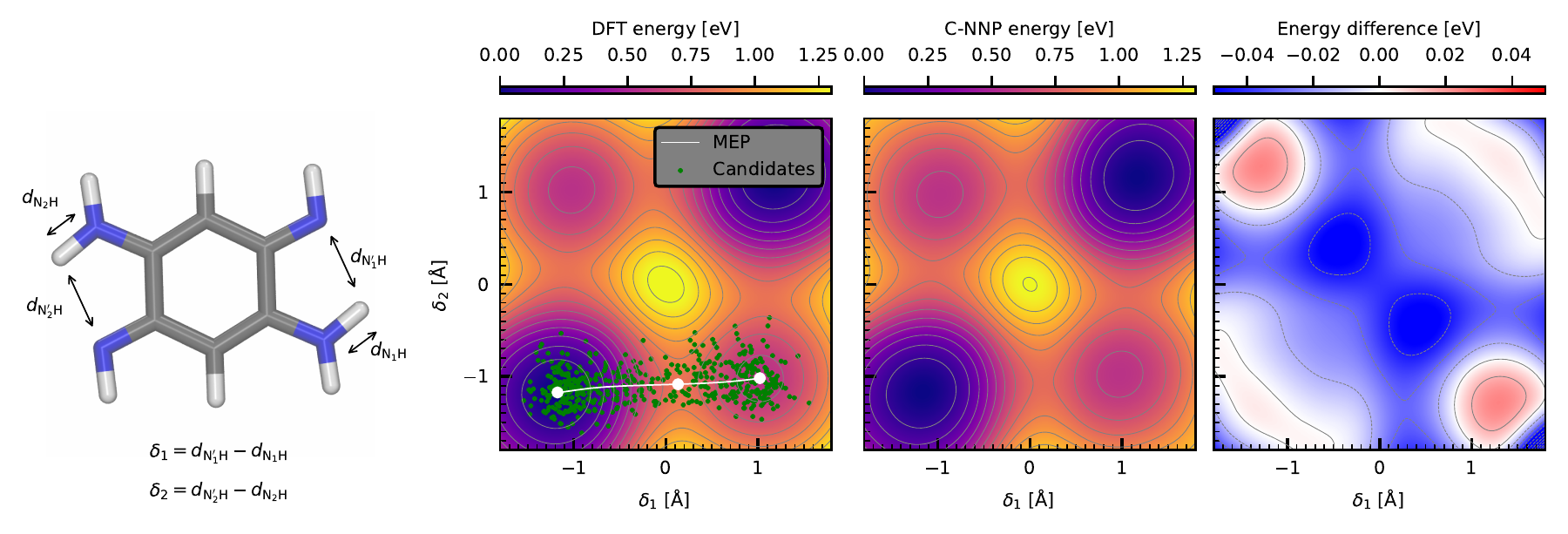}
    \caption{
        Comparison between the reference \textit{ab initio} revPBE0-D3 and the C-NNP proton-sharing PESs of the DABQDI molecule.
        The left panel shows the projection is the reference DFT PES into the $\delta_1$, $\delta_2$ subspace using a color scale as well as individual isoenergetic contours.
        Furthermore, the minimal non-trivial MEP which describes a single proton transfer is depicted in white with the selected control points highlighted.
        A sparse subset of the TTS geometries sampled around the MEP at \SI{300}{K} is shown in green.
        The middle panel shows the corresponding PES projection calculated with the resulting C-NNP.
        Finally, the right panel shows the difference between the two PESs and quantifies the maximum error of the C-NNP fit in the classically thermally accessible regions on the order of 10$^{-3}$ to 10$^{-2}$~eV.
        The two $\delta$ coordinates are illustrated in the snapshot of the DABQDI molecule to the left of the panels.
    }
    \label{fig:dabqdi-pes}
\end{figure*}

The corresponding PES reduced to the relevant $\delta_1$ and $\delta_2$ subspace (illustrated in the snapshot in Figure~\ref{fig:dabqdi-pes}), where each proton-sharing coordinate describes a single proton-sharing site, was obtained at the revPBE0-D3 level of electronic structure theory through a relaxed scan of the molecular potential energies while applying appropriate constraints and is shown in the left panel of Figure~\ref{fig:dabqdi-pes}.
The shown data was aligned so that the global minimum of the PES corresponds to the zero-energy level.
The typical structure of the PES featuring four distinct configurational minima and four transition states corresponds to a sequential double proton transfer at the level of an MEP.
Here, one proton is first fully exchanged to reach an intermediate state located at a higher potential energy and only then the other proton follows.
The height of the potential energy barrier for this sequential process of roughly \SI{0.8}{eV} indicates that its thermal rate should be negligible.
The alternative concerted proton transfer path that is seen in other species including carboxylic acid dimers~\cite{Smedarchina2008/10.1524/zpch.2008.5389} is classically disallowed in this case by a tall ($>$\SI{1.2}{eV}) potential barrier in the middle of the presented PES which represents a second-order saddle point and, as such, no MEP can go through it.
The symmetry of the DABQDI PES allows us to explicitly address only a single proton transfer: unlike in the previous example, we exploit this feature here for the C-NNP model generation.
The relevant non-trivial MEP connecting the two chemically distinct minima was obtained using the CI-NEB optimization at the revPBE0-D3 level of theory and is shown in white in the left panel of Figure~\ref{fig:dabqdi-pes}.
From there, the straightforward TTS candidate generation was performed using the three usual control points in the two minima and in the transition state at \SI{300}{K} with the linear sampling density of \SI{1e3}{\angstrom ^ {-1}}.
A subset of the candidate geometries is shown in the left panel of Figure~\ref{fig:dabqdi-pes} as green points.
The obtained C-NNP model was used to recreate the 2D proton transfer PES which is shown in the middle panel of Figure~\ref{fig:dabqdi-pes} with the energies aligned in the same way as in the DFT case.
Qualitatively, the C-NNP model captures all the features of the original \textit{ab initio} PES including the position of the minima, the transition state, and the barriers, as well as the potential energy values.
Note that the good agreement in the representation of the central barrier in spite of the lack of corresponding geometries in the TTS candidates is due to the successful extrapolation by the model.
The quantitative difference between the \textit{ab initio} and C-NNP potential energy landscape is captured in the right panel of Figure~\ref{fig:dabqdi-pes} and shows that the potential energies are reproduced as closely as approximately $\pm$\SI{0.05}{eV}.
This difference could be decreased further, if desired, by changing the architecture of the C-NNP and by increasing the size of the training set beyond the current 620 structures.

\begin{figure}[tb!]
    \centering
    \includegraphics{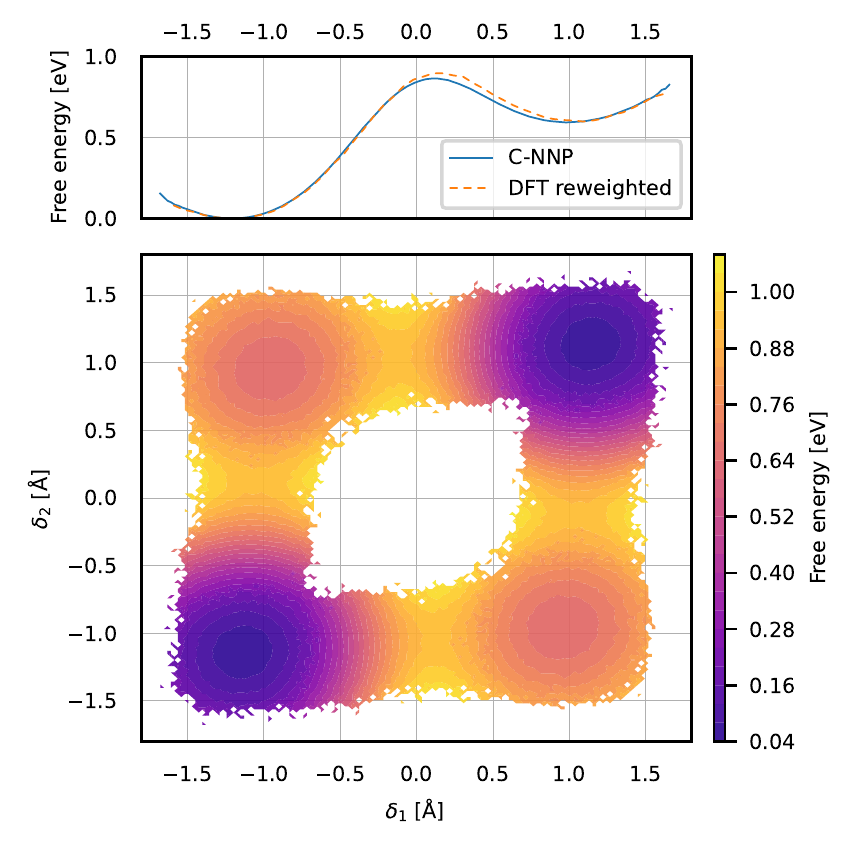}
    \caption{
        Umbrella sampling simulation of the single proton transfer in the DABQDI molecule along the $\delta_1$ collective variable using the C-NNP potential.
        The top panel shows the obtained free energy profile in blue.
        For validation purposes, we also show the DFT free energy profile (orange, dashed) obtained by reweighting the C-NNP configurations as described in the text.
        Note that no umbrella sampling using the DFT potential was performed to obtain the DFT free energy profile.
        The bottom panel shows the full 2D free energy surface obtained by weighing the distribution in the two proton-sharing coordinates by the thermal Boltzmann factors extracted from the biased simulation and symmetrizing the resulting histogram.
    }
    \label{fig:dabqdi-fes}
\end{figure}

Since DABQDI features barriers that are not practically accessible by direct MD, it serves as a useful example to illustrate the power of the TTS-based C-NNP model to perform an enhanced sampling calculation to correctly estimate the free energy profile of the double proton transfer at \SI{300}{K}.
This was obtained using an umbrella sampling simulation in the coordinate $\delta_1$ with the C-NNP PES (as described in Section~\ref{sec:computational-details}) followed by a multistate Bennet acceptance ratio (MBAR) reweighing of the biased configurations.
The obtained 1D free energy profile in $\delta_1$ is shown in blue in the top panel of Figure~\ref{fig:dabqdi-fes}.
The transition state is located at roughly \SI{0.8}{eV} above the global minimum.
Comparing this with the value of the corresponding potential energy suggests that the entropic contribution in the gas phase system is small and that the population at the barrier is clearly negligible at \SI{300}{K}.
To validate the obtained free energy profile, we perform a reweighing of a subset of the obtained configurations in each umbrella window to the original DFT ensemble.
This is achieved by additionally multiplying each MBAR-obtained unbiased weight by the corresponding factor $e^{-\beta\Delta E}$ where the energy difference $\Delta E$ is the difference between the C-NNP and DFT potential energy for each configuration.
For this purpose, we used a total of 3000 configurations obtained by selecting 100 geometries evenly spaced in time from each of the 30 umbrella sampling windows.
The resulting profile, which is a good approximation to the full-DFT free energy profile, is displayed as the orange dashed line in Figure~\ref{fig:dabqdi-fes} and shows very good correspondence with the profile obtained using the C-NNP model alone.
This procedure thus at the same time validates the C-NNP model and provides DFT data for a fraction of the cost of the hypothetical purely \textit{ab initio} enhanced sampling simulation.
Monitoring the values of the collective variable $\delta_2$ along the umbrella sampling simulation and using the thermal weights obtained from the MBAR treatment of the biased simulations also allows recovering the 2D free energy surface in $\delta_1$ and $\delta_2$.
Its symmetrized version is shown in the bottom panel of Figure~\ref{fig:dabqdi-fes}.
It is worth noting that the computational acceleration of the C-NNP umbrella simulation in comparison to the naive execution with the original DFT method is substantial.
To illustrate the computational savings, we can compare the times required for one MD step with the implementations in CP2K used in this work.
With the hybrid functional, one MD step takes \SI{272}{s} on a single core or \SI{17}{s} on 32 cores (a full node) of our EPYC-based cluster.
With the C-NNP, one step takes \SI{0.006}{s} and does not scale meaningfully to more cores due to the small system size.
This yields a speedup of $\sim$45000$\times$ on identical resources or $\sim$2800$\times$ with more resources given to the DFT calculation.
Obviously, the specific numbers will depend on the details of the electronic structure setup and the MLP architecture used, as well as the specific implementations and hardware used, but this behavior of our particular setup should provide a general idea.

\section{Conclusions}
\label{sec:conclusions}

In this work, we have introduced the TTS method to sample thermal geometries around MEPs that describe barrier-crossing transitions in molecular systems. 
The goal of the method is to provide a physically meaningful set of candidate structures for the creation of MLPs without the need to run computationally demanding \textit{ab initio} simulations.
In our case specifically, we submit these geometries to QbC and construct a C-NNP model, but the same candidates could be used for other types of models as well.
The execution of the TTS protocol as a whole entails a relatively modest computational cost with respect to the original \textit{ab initio} method that is given by the MEP optimization, several Hessian evaluations, and a small number of single-point \textit{ab initio} calculations for the QbC-selected geometries.
In terms of application to realistic systems, the TTS method yields highly accurate C-NNP models in all studied cases.
This was achieved either by using the generated candidate set directly, or by letting the resulting C-NNP model undergo additional active learning generations to compensate for a pronounced anharmonic effect as seen in the case of malonaldehyde.
As such, the performance of TTS in the presented test systems demonstrates its robustness and efficiency and suggests applicability in most gas-phase systems, including highly anharmonic cases.

A noteworthy feature of the TTS method is its ability to provide thermal geometries sampled from the quantum thermal distribution at essentially the computational cost of the classical case.
As such, models that are appropriate for use in path integral simulations are made readily available without the need to run expensive PI-AIMD simulations at all.
However, it is important to recognize that although the present formulation of TTS can address quantum behavior, it has limitations in this regard that derive from the fundamentally classical nature of the MEP.
Nuclear quantum effects, in particular quantum tunneling through the potential barrier, can cause the configuration-space probability density of the system to deviate from the transition tube around the MEP in a way that renders the coverage by TTS samples insufficient.

To account for this, the above formulation of TTS can be straightforwardly generalized from sampling around classical MEPs to ring-polymer instantons~\cite{Kaestner2014/10.1002/wcms.1165}, which represent the paths of optimal tunneling.
While this modification requires essentially no adaptation of the TTS theory and implementation itself, one can anticipate an elevated computational cost due to the required instanton optimization at the explicit \textit{ab initio} level. 
The approach will find applications beyond the gas phase, in systems where vibrational normal modes are a meaningful concept, such as crystals and surface-bound molecules.
Since our research anticipates the need to address such systems in the near future, we expect TTS to be a valuable tool in the creation of accurate, yet computationally accessible potentials that will enable the accurate description of these more complex systems at unprecedented sizes and simulation time scales.

\begin{acknowledgments}

K.B. and H.B. acknowledge funding from grant schemes at Charles University, reg. n. CZ.02.2.69/0.0/0.0/19\_073/0016935, and from the IMPRS for Quantum Dynamics and Control.
O.M. acknowledges support from the Czech Science Foundation, project No. 21-27987S.

\end{acknowledgments}

\section*{References}

%
\end{bibunit}


\clearpage

\setcounter{section}{0}
\setcounter{equation}{0}
\setcounter{figure}{0}
\setcounter{table}{0}
\setcounter{page}{1}

\renewcommand{\thesection}{S\arabic{section}}
\renewcommand{\theequation}{S\arabic{equation}}
\renewcommand{\thefigure}{S\arabic{figure}}
\renewcommand{\thetable}{S\arabic{table}}
\renewcommand{\thepage}{S\arabic{page}}
\renewcommand{\citenumfont}[1]{S#1}
\renewcommand{\bibnumfmt}[1]{$^{\rm{S#1}}$}

\title{Supporting information for: \mytitle}
{\maketitle}

\begin{bibunit}

\nocite{revtex-control}

\section{Additional Theory Details\label{si-sec:theory}}

The following section contains additional details on the theory discussed in the main text.
First, the derivation of the effective quantum temperature of the quantum harmonic oscillator is presented.
This is followed by the discussion of the density-matching condition needed in the TTS procedure for appending thermal NMS of the minima away from MEPs.

\subsection{Quantum effective temperature}

To derive the expression for the quantum effective temperature, we start from the standard Hamiltonian for a 1D harmonic oscillator in a mass-weighted coordinate $\Omega$ with the natural frequency $\omega$
\begin{equation}
    \widehat{H}
    =
    -\frac{\hbar^2}{2}\frac{\dd^2}{\dd \Omega^2} + \frac{1}{2}\omega^2\Omega^2.
\end{equation}

The $n$-th solution for the corresponding time-independent Schr\"{o}dinger equation ($n=0,1,2,\dots,\infty$) is in the well-known form based on the physicist's Hermite polynomials $H_n$
\begin{equation}
    \psi_n(\Omega)
    =
    \left( \frac{1}{2^n n! \sqrt{\pi}} \right)^\frac{1}{2}
    \left( \frac{\omega}{\hbar} \right)^\frac{1}{4}
    e^{-\frac{\omega \Omega^2}{2\hbar}}
    H_n\left( \sqrt{\frac{\omega}{\hbar}} \Omega \right),
\end{equation}
which can be simplified to
\begin{equation}
    \psi_n(y)
    =
     \left( \frac{\alpha}{2^n n! \sqrt{\pi}} \right)^\frac{1}{2}
     e^{-\frac{y^2}{2}} H_n(y)
\end{equation}
with the substitutions $\alpha = \sqrt{\frac{\omega}{\hbar}}$ and $y = \alpha\Omega$.
The corresponding energy levels are
\begin{equation}
    E_n
    =
    \hbar\omega\left( n + \frac{1}{2} \right).
\end{equation}

Ultimately, we are interested in the thermal density of the harmonic oscillator $\rho(y)$, which is formally obtained as the diagonal elements of the full density matrix $\rho(y, y')$.
A formula for the density matrix in terms of an infinite sum over all states can be obtained as
\begin{equation}
\begin{split}
    \rho(y, y')
    &=
    \bra{y'} e^{-\beta \widehat{H}} \ket{y} \\
    &=
    \sum_{n, m} \braket{y'}{\psi_n}\bra{\psi_n} e^{-\beta \widehat{H}} \ket{\psi_m} \braket{\psi_m}{y} \\
    &=
    \sum_{n} \psi_n^*(y') \psi_n(y) e^{-\beta E_n } \\
    &=
    e^{-\frac{\beta\hbar\omega}{2}} \frac{\alpha}{\sqrt{\pi}}
    \sum_{n} \frac{1}{2^n n!} e^{ -\beta \hbar \omega n } \cdot \\
    &\cdot e^{-\frac{y^2 + y'^2}{2}} H_n(y) H_n(y'),
\end{split}
\end{equation}
which can be closed using the so-called Mehler kernel~\cite{Mehler1886}
\begin{equation}
    \sum_n \frac{\chi^n}{2^n n!} H_n(x) H_n(y)
    =
    \frac{1}{\sqrt{1 - \chi^2}} e^{-\frac{\chi^2(x^2 + y^2) - 2\chi xy}{1 - \chi^2}}:
\end{equation}
a mathematical identity for a parameter $\chi$ that allows summing over the above product of Hermite polynomials.
With this, one obtains
\begin{equation}
\begin{split}
    \rho(y, y')
    &=
    e^{-\frac{\beta\hbar\omega}{2}} \frac{\alpha}{\sqrt{\pi}} e^{-\frac{y^2 + y'^2}{2}} 
    \frac{1}{\sqrt{1 - e^{-2\beta \hbar \omega }}} \cdot \\
    &\cdot e^{-\frac{e^{-2\beta\hbar\omega(y^2 + y'^2) - 2e^{-\beta\hbar\omega}yy'}}{1 - e^{-2\beta\hbar\omega}}}.
\end{split}
\end{equation}
The thermal density is then simply obtained by setting $y=y'$, which gives
\begin{equation}
\begin{split}
    \rho(y)
    &=
    e^{-\frac{\beta\hbar\omega}{2}} \frac{\alpha}{\sqrt{\pi}}
    \frac{1}{\sqrt{1 - e^{-2\beta \hbar \omega }}} \cdot \\
    &\cdot e^{-y^2 \left[ 1 + \frac{2e^{-\beta\hbar\omega}\left( e^{-\beta\hbar\omega} - 1 \right)}{1 - e^{-2\beta\hbar\omega}} \right]} \label{si-eq:mehler}
\end{split}
\end{equation}
after trivial rearrangements.
As discussed in the main text, the thermal density is Gaussian with an $\omega$-dependent normalization factor (first line of Equation~\ref{si-eq:mehler}) and a non-trivial scaling of the exponent. 
The latter will be rearranged to extract the quantum effective temperature from Equation~\ref{eq:betastar} of the main text.
Substituting $t = e^{-\beta \hbar \omega}$, we get
\begin{equation}
\begin{split}
    & 1 + \frac{2e^{-\beta\hbar\omega}\left( e^{-\beta\hbar\omega} - 1 \right)}{1 - e^{-2\beta\hbar\omega}}
    =
    1 + \frac{2t(t-1)}{1-t^2} \\
    &=
    \frac{t^2 - 2t + 1}{1 - t^2}
    =
    \frac{1 - t}{1 + t}
    =
    \frac{1 - e^{-\beta \hbar \omega}}{1 + e^{-\beta \hbar \omega}} \\
    &=
    \tanh\left(\frac{\beta \hbar \omega}{2}\right),
\end{split}
\end{equation}
which gives
\begin{equation}
    \rho(\Omega)
    \propto
    e^{-\frac{\omega}{\hbar} \tanh\left( \frac{\beta \hbar \omega}{2}\right) \Omega^2}
\end{equation}
for the thermal density in the original coordinates.
Comparing the hyperbolic tangent factor with the form of a classical distribution ($\rho_{\mathrm{cl}}(y) \propto e^{-\beta V} = e^{-\frac{1}{2}\beta \omega^2 \Omega^2}$) of the harmonic oscillator
at a new inverse temperature $\beta^*$, we immediately get
\begin{equation}
\begin{split}
    &\frac{\omega}{\hbar} \tanh\left(\frac{\beta\hbar \omega}{2}\right)
    \equiv
    \frac{1}{2}\beta^* \omega^2 \\
    &
    \beta^*(\beta, \omega)
    = \frac{2}{\hbar \omega} \tanh\left( \frac{\beta \hbar \omega}{2}\right), 
\end{split}
\end{equation}
as stated in the main text.
Since the hyperbolic tangent approaches unity as its argument grows, it is always lower-valued than a linear function and thus always $\beta^* < \beta$. 
In turn, this means that the quantum harmonic oscillator is always at an effective temperature higher than the reference classical temperature and its density is thus always wider.
Moreover, note that the classical limit $\beta^* \rightarrow \beta$ is reached as $\beta \rightarrow 0$ or $\omega \rightarrow 0$: at a given temperature, the higher frequency modes will have a more pronounced quantum effect and, in turn, at a given frequency, the quantum effect will be more pronounced at a lower temperature.

\subsection{Density matching}

An obvious technical problem with appending the thermal NMS at the minima at the MEP endpoints is the matching of the sampling density so that it is a smooth continuation of the linear MEP sampling (as indicated in Figure~\ref{fig:wurst} of the main text).
For practical purposes, one can achieve sufficient smoothness by separately choosing a linear density for the MEP sampling and a fixed number of points that are sampled thermally around the minima by a trial-and-error approach.
However, this does not lead to an analytically smooth transition.
That can be achieved by integrating the thermal Gaussian distributions at the minima over all dimensions but the one defined by the tangent vector to the MEP  $\boldsymbol{\tau}$ at the minimum and finding a scaling factor that brings the density at the maximum of the Gaussian equal to the linear density $\rho_0$ of sampling along the MEP, yielding an analytic continuation.
Rather inconveniently from a mathematical perspective, the MEP tangent $\boldsymbol{\tau}$ is, in general, not identifiable with any of the normal modes $\boldsymbol{\Omega}_i$, which diagonalize the Gaussian distribution.
Therefore, to perform the integration over everything but the $\tau$-direction, one needs to set up a new orthogonal basis built around this vector in which, however, the Gaussian is no longer diagonal.
Still, the result of such integration will be a 1D Gaussian and its width will be given by the effective frequency $\widetilde{\omega}$
\begin{equation}
    \rho(\tau)
    =
    \int \rho(\tau, \tau_2, \dots, \tau_{3N}) \ \dd\tau_2 \dots \dd\tau_{3N}
    \propto
    e^{-\frac{1}{2}\beta\widetilde{\omega}^2 \tau^2}.
\end{equation}
It can be shown (with the full proof being outside of the scope of the present work), that this effective frequency can be calculated from the following compact formula
\begin{equation}
    \widetilde{\omega}^2
    =
    \frac{\det (\mathbb{H})}{\det_{\tau, \tau} (\mathbb{H})},
\end{equation}
where $\mathbb{H}$ is the Hessian matrix of the potential at the minimum and the symbol $\det_{\tau, \tau}$ denotes the minor of the Hessian matrix taken over the row and the column corresponding to the $\tau$-dimension.
Note that, unlike the full determinant, the minor is a basis-dependent operation and, therefore, the Hessian matrix must be transformed to the $\tau$-oriented basis prior to the calculation of $\widetilde{\omega}$.
The effective standard deviation of the 1D distribution is $\widetilde{\sigma} = 1 / \sqrt{\beta\widetilde{\omega}^2}$.
The aim at this point is to find a scaling factor $\alpha$ which ensures that the density at the configurational minimum $\alpha\rho(0)$ matches $\rho_0$.
However, one cannot simply equate the two values directly, since the $\rho(\tau)$ comes from a mass-weighted coordinate system (note that the $\boldsymbol{\tau}$-based basis was obtained as a simple rotation of the normal mode basis, which itself is naturally mass-weighted) and $\rho_0$ is understood as a density along the non-weighted Cartesian MEP curve.
Therefore, we need to calculate the effective mass $\mu$ of the $\tau$-direction, which is the following weighted average
\begin{equation}
    \mu
    =
    \frac{\boldsymbol{\tau}^T \mathbb{M} \boldsymbol{\tau}}{\boldsymbol{\tau}^T \boldsymbol{\tau}},
\end{equation}
where $\mathbb{M}$ is the diagonal mass matrix.
Then, we obtain the analytic expression
\begin{equation}
    \alpha \rho(0) 
    =
    \frac{\alpha}{\widetilde{\sigma} \sqrt{2\pi}}
    = 
    \sqrt{\frac{1}{\mu}} \rho_0
\end{equation}
which immediately gives the crucial density-matching condition
\begin{equation}
    \alpha
    =
    \widetilde{\sigma} \sqrt{\frac{2\pi}{\mu}} \rho_0.
\end{equation}
The scaling parameter $\alpha$ thus captures how many times the original normalized density $\rho(\tau)$ needs to be increased in order to match at its peak the sampling density at the MEP.
In practice, where the sampling of configurations is discrete and finite, one needs to sample the nearest-integer-to-$\alpha$ points to match the sampling density at the MEP.

\section{Additional results}
\label{si-sec:additional-results}

\subsection{TTS of malonaldehyde proton transfer}

\begin{figure}[htb!]
    \centering
    \includegraphics[width=\linewidth]{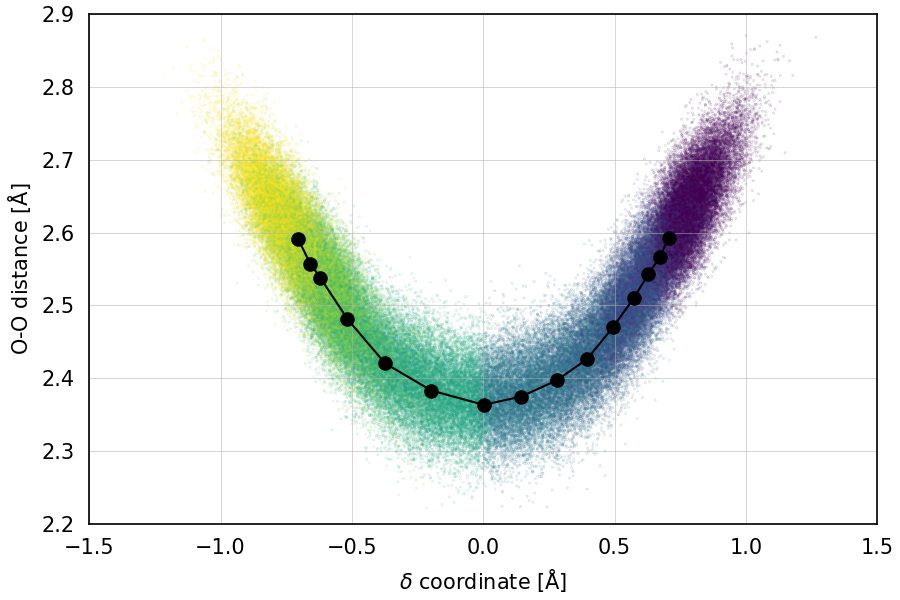}
    \includegraphics[width=\linewidth]{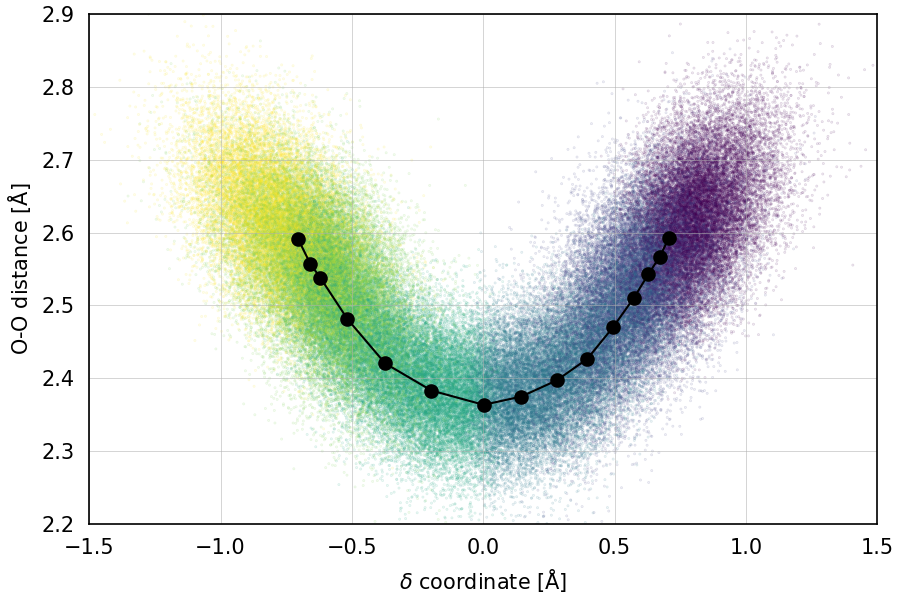}
    \caption{
        Thermal geometries along the malonaldehyde proton-sharing MEP at \SI{300}{K} generated using the classical formulation of TTS (top panel) and the quantum formulation of TTS that relies on $\beta^*$ (bottom panel).
    }
    \label{fig-si:malonaldehyde-TTS}
\end{figure}

The top panel of Figure~\ref{fig-si:malonaldehyde-TTS} shows the distribution of the TTS geometries of malonaldehyde obtained at \SI{300}{K} using the classical harmonic distributions in the mode directions perpendicular to the proton-sharing MEP.
Three control points were selected for the TTS: the two symmetric minima at the edges of the MEP and the transition state at $\delta =$ \SI{0}{\angstrom}.
Each candidate geometry is colored by the assignment to its control point following the methodology described in Section~\ref{sec:theory} of the main text and illustrated in Figure~\ref{fig:wurst} therein.
The bottom panel of Figure~\ref{fig-si:malonaldehyde-TTS} then shows the same situation but using the quantum formulation of TTS in which $\beta^*$ is calculated for each degree of freedom.
The use of the quantum densities leads to the characteristic broadening of the distribution of the geometries.

\subsection{Path integral MD for malonaldehyde}
\label{si-sec:pimd-for-malonaldehyde}

\begin{figure*}[tb!]
    \centering
    \includegraphics{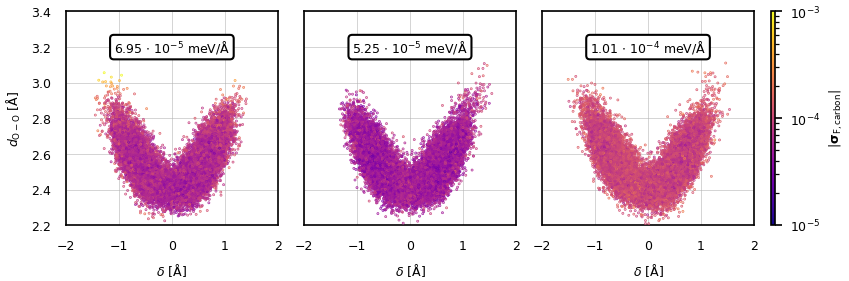}
    \caption{The distribution of force disagreement of the carbon atoms along a PIMD trajectory for three different MLPs.
    The initial model trained on structures selected from the TTS candidate set (generated at an effective temperature of \SI{300}{K} for path integral structures) is shown on the left and the model improved with structures selected from the PIMD trajectory of the initial model is shown in the middle panel.
    The model used to generate the data in the right panel was trained only on TTS structures, combining the contributions from the proton transfer MEP and the MEP of the torsion along the C--C single bond.
    The force disagreement is indicated by the color scale on the right and the mean disagreement in noted in each panel.
    While the middle and right panels exhibit an even distribution of disagreement, there is a sharp increase in disagreement at the tails of the distribution in the left panel.
    In all cases, only a subset of the complete PIMD trajectory is shown.}
    \label{si-fig:malonaldehyde2q}
\end{figure*}

The process described in the main paper for classical MD simulation can be done analogously for PIMD.
TTS was used to generate structures along the proton-sharing MEP at an effective temperature of \SI{300}{K} and 620 structures were selected during one QbC run for the initial model.
Similar to the classical model, force disagreement during PIMD was low around the two free energy minima and along the MEP, but increased at the tails of the distribution, as can be seen in the left panel of Figure~\ref{si-fig:malonaldehyde2q}
To improve the MLP in these areas of the configuration space, two strategies were pursued.
An iterative process, in which PIMD simulations and QbC processes with the structures of the PIMD trajectory as candidates are alternated until force disagreement is low along the whole PIMD trajectory.
Or a single QbC run, in which TTS is used to create candidate structures along the MEP of the proton-transfer reaction as well as the MEP of the torsion of the C--C single bond of the propane back bone up until the transition state to include the strongly anharmonic vibrations in this direction of the configuration space.
The distribution of the force disagreement for the second-generation MLP of the iterative process can be seen in the middle panel of Figure~\ref{si-fig:malonaldehyde2q}.
Unlike the classical case in the main paper, a single additional PIMD--QbC sequence, which added 435 structures to the training data set, was sufficient to obtain an MLP capable of accurate PIMD simulations.
In the right panel of Figure~\ref{si-fig:malonaldehyde2q}, the results from the extended TTS MLP can be seen.
As for the classical case, the resulting model gives a low committee disagreement along the whole PIMD trajectory.

\subsection{Validation of malonaldehyde MLPs}
\label{si-sec:validation-of-malonaldehyde}

\begin{table}[htb!]
\caption{Test root mean square errors for energies and forces. IA stands for iterative approach.}
\begin{tabular}{lrr}
\hline
MLP                    & RMSE E [meV]          & RMSE F [meV $\mathrm{\AA^{-1}}$] \\ \hline
IA generation 1        & 24.58                 & 161.7                                            \\
IA generation 2        & 1.89                  & 18.7                                             \\
IA generation 3        & 1.80                  & 18.3                                             \\
extended TTS           & 3.44                  & 24.4                                             \\ \hline
\end{tabular}
\label{tab-si:malonaldehyde-validation}
\end{table}

\begin{figure}[htb!]
    \centering
    \includegraphics[width=\linewidth]{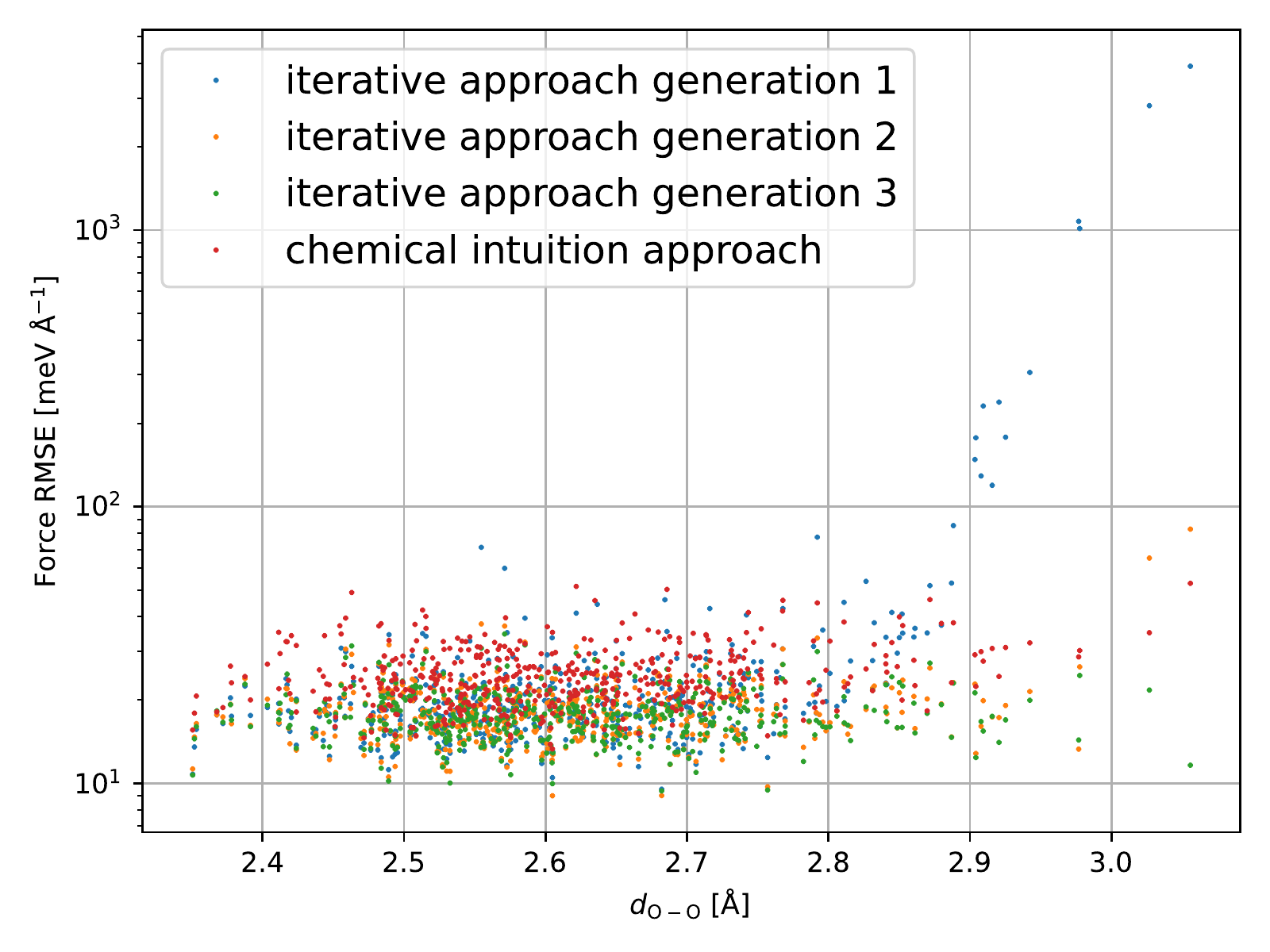}
    \caption{
        Distribution of force RMSEs for the three generations of MLPs from the iterative approach and the MLP from the extended TTS approach along the distance between the two oxygen atoms. Of the 1000 structures of the test set, only the 500 structures sampled from an MD trajectory are shown here for clarity, as this subset includes the most insightful structures.
    }
    \label{fig-si:malondaldehyde-validation}
\end{figure}

In the main text of the paper, the energy and force RMSEs on an independent test set for the final models were reported.
The complete set of validation errors can be seen in table \ref{tab-si:malonaldehyde-validation}.
While the later generations of the iterative approach and the extended TTS model performed well, the initial model performed poorly.
However, the errors averaged over the whole test set tell only a part of the story, because they are not evenly distributed for all MLPs.
As shown in Figure \ref{fig-si:malondaldehyde-validation}, the first generation model displays low force errors for most structures with a low $d_{\mathrm{O-O}}$, which accounts for structures along the proton-sharing transition and the minima, but exceedingly large errors of up to \SI{603}{meV} (energy) and \SI{3908}{meV \angstrom ^ {-1}} (forces) for $d_{\mathrm{O-O}} > 2.8$.
Due to the anharmonicity of the system, no structures from this region of the configuration space are included in the TTS along the proton sharing coordinate and hence such structures are absent from the training data set.
Therefore, the model is not suitable for accurate MD simulations.
The same effect can be observed in the second generation model, however less severe.
The final models of both approaches, on the other hand, exhibit an even distribution of force RMSEs across the complete span of oxygen-oxygen distances.

\subsection{Overlap of umbrella windows in DABQDI}

\begin{figure}[htb!]
    \centering
    \includegraphics[width=\linewidth]{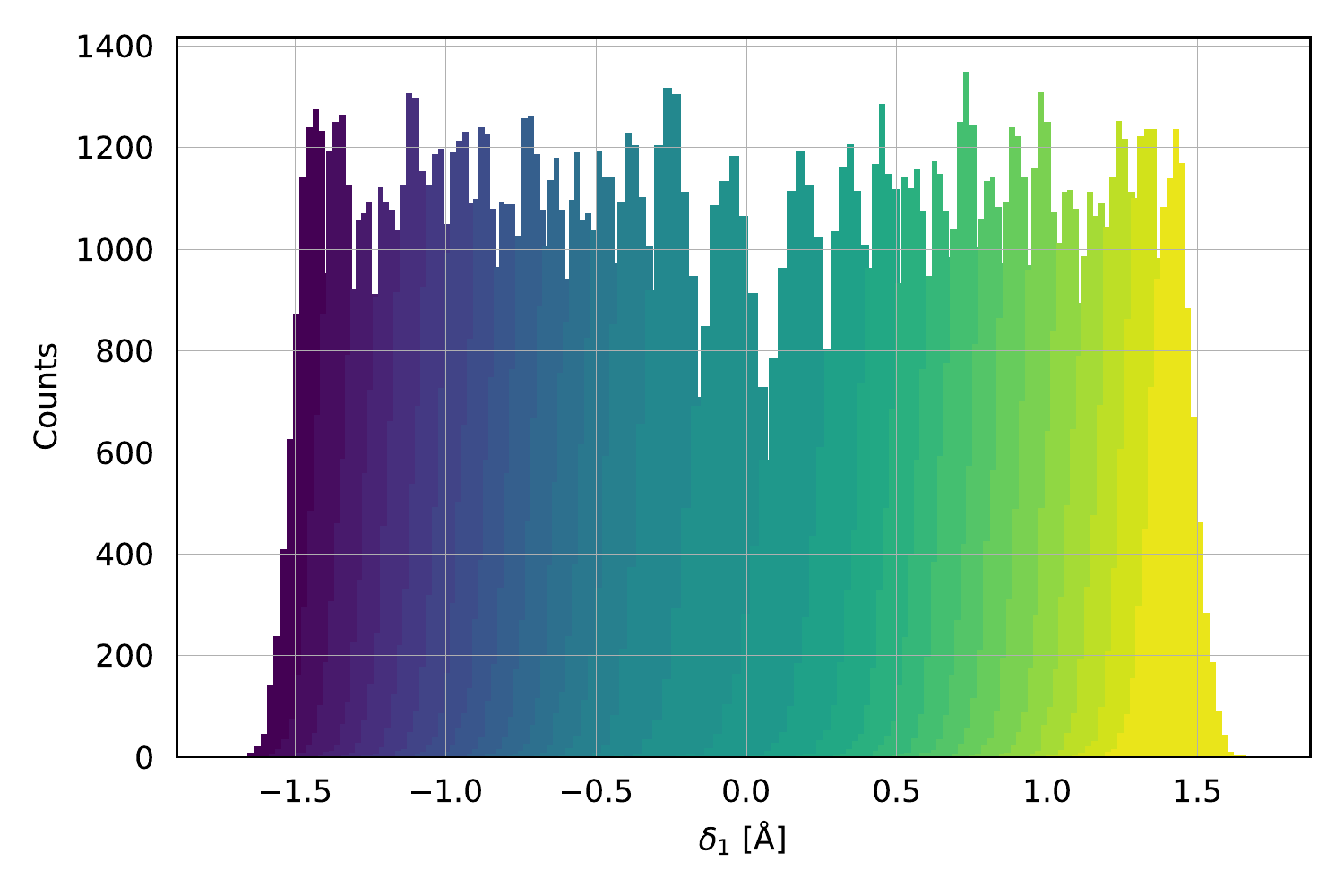}
    \caption{Histograms of $\delta_1$ values observed in each simulation window during the course of the umbrella sampling simulation of the DABQDI proton transfer reaction.}
    \label{si-fig:histograms}
\end{figure}

Figure~\ref{si-fig:histograms} shows that the histograms of $\delta_1$ values from the individual windows of the umbrella sampling simulation of the proton transfer in the DABQDI molecule exhibit sufficient overlap.
Therefore, the combination of window spacing and harmonic restraint stiffness selected to obtain the results shown in Figure~\ref{fig:dabqdi-fes} of the main text represents an appropriate choice.

\section*{References}


%
\end{bibunit}

\end{document}